\documentclass{article}

\usepackage[hidelinks]{hyperref}
\usepackage{ml4cps}
\usepackage[T1]{fontenc}
\usepackage{amsfonts}
\usepackage{url}
\usepackage{booktabs}
\usepackage{graphicx}
\usepackage{doi}
\usepackage{amsmath}
\usepackage{amsthm}
\usepackage{cleveref}
\usepackage{subfig}
\usepackage{float}
\usepackage{colortbl}   
\usepackage{xcolor}  
\usepackage{bm}  
\newtheorem{theorem}{Theorem}
\usepackage{amsthm}

\usepackage{titlesec}
\usepackage{color}

\usepackage[utf8]{inputenc}
\usepackage[english]{babel}
\usepackage[T1]{fontenc} % Font encoding
\usepackage{amsfonts}

\usepackage{graphicx}
\graphicspath{{Images/}}
\usepackage{eso-pic} % For the background picture on the title page
\usepackage{subfig} % Numbered and caption subfigures using \subfloat
\usepackage{caption} % Coloured captions
\usepackage{transparent}

\usepackage{amsmath}
\usepackage{amsthm}
\usepackage{bm}
\usepackage[overload]{empheq}  % For braced-style systems of equations

\usepackage{tabularx}
\usepackage{longtable} % tables that can span several pages
\usepackage{colortbl}

\usepackage{algorithm}
\usepackage{algorithmic}

\usepackage{natbib} % Square brackets, citing references with numbers, citations sorted by appearance in the text and compressed
\usepackage{appendix}

\usepackage{enumitem}

\usepackage{amsthm,thmtools,xcolor} % Coloured "Theorem"
\usepackage{comment} % Comment part of code
\usepackage{fancyhdr} % Fancy headers and footers
\usepackage{lipsum} % Insert dummy text
\usepackage{tcolorbox} % Create coloured boxes (e.g. the one for the key-words)

\definecolor{lavender}{rgb}{0.9, 0.9, 0.98}

\newcommand{\bea}{\begin{eqnarray}} % Shortcut for equation arrays
\newcommand{\eea}{\end{eqnarray}}
\title{Calibrated quantile prediction %models 
for Growth-at-Risk}

\author{%
    \begin{tabular}{c c}
        \normalsize Pietro Bogani & \normalsize Matteo Fontana \\
        \normalfont \footnotesize Politecnico di Milano & \normalfont \footnotesize Royal Holloway, University of London \\[1em]
        \normalsize Luca Neri & \normalsize Simone Vantini \\
        \normalfont \footnotesize CORE, UCLouvain & \normalfont \footnotesize Politecnico di Milano \\
    \end{tabular}%
}

\renewcommand{\shorttitle}{Calibrated quantile prediction %models 
for Growth-at-Risk}

\hypersetup{
pdftitle={Calibrated quantile prediction %models 
for Growth-at-Risk},
pdfauthor={Pietro Bogani, Matteo Fontana, Luca Neri, Simone Vantini},
pdfkeywords={Conformal Prediction, Quantile Estimation, Growth-at-Risk, Quantile Regression, Risk Management}
}

\begin{document}

\maketitle

\begin{abstract}
Accurate computation of robust estimates for extremal quantiles of empirical distributions is an essential task for a wide range of applicative fields, including economic policymaking and the financial industry. Such estimates are particularly critical in calculating risk measures, such as Growth-at-Risk (GaR). % and Value-at-Risk (VaR). 
This work proposes a conformal framework to estimate calibrated quantiles, and presents an extensive simulation study and a real-world analysis of GaR to examine its benefits with respect to the state of the art.
Our findings show that CP methods consistently improve the calibration and robustness of quantile estimates at all levels. The calibration gains are appreciated especially at extremal quantiles, which are critical for risk assessment and where traditional methods tend to fall short. In addition, we introduce a novel property that guarantees coverage under the exchangeability assumption, providing a valuable tool for managing risks by quantifying and controlling the likelihood of future extreme observations.
\end{abstract}


\section{Introduction}
Policymakers and financial institutions increasingly rely on Growth-at-Risk (GaR) to quantify downside risks to economic growth, yet existing methods often fail to deliver well-calibrated estimates of tail risks. A well-calibrated quantile accurately reflects empirical coverage, which means that the $\tau$ -th quantile of a series, for $\tau \in [0,1]$, should have $\tau\times 100 \%$ of the data points falling below its value.  While GaR has become a standard tool for macroeconomic risk assessment since \cite{adrian_vulnerable_2019}, accurately estimating the distribution of future growth outcomes remains challenging, particularly in the presence of non-linearities and structural breaks.
By assessing the likelihood of adverse economic conditions and their potential impact, GaR provides a more comprehensive understanding of economic vulnerability and resilience. 

Since its introduction in \cite{adrian_vulnerable_2019}, GaR has become a popular tool used to inform policy makers \citep[see, e.g.,][]{international_monetary_fund_global_2017}. 
Central banks and financial regulators use GaR to assess systemic risks and inform monetary policy decisions. Investment managers leverage these estimates for portfolio optimisation, while government agencies rely on them to anticipate fiscal impacts of economic shocks. However, the effectiveness of these applications depends crucially on having well-calibrated estimates of the full growth distribution - particularly in the tails where standard methods often fail. 

In GaR analysis, non-linearities and asymmetries acquire a central role for understanding the propagation mechanisms of adverse financial conditions onto real growth. The importance of requiring methods that include such features has been magnified in recent years as a result of several converging factors. Firstly, the global economy has faced increased uncertainty and volatility, driven by events such as the COVID-19 pandemic, geopolitical tensions, and climate-related issues. These events have underscored the need for robust and flexible risk assessment tools that can capture asymmetries and tail risks to identify systemic vulnerabilities that traditional forecasting models, based on central tendency forecasts (means or medians), may overlook.

% APPROACHES TO GaR
% Central banks and policymakers can use GaR to better assess economic risks and make informed decisions about monetary policy, while financial regulators can leverage it to monitor systemic risks and improve financial stability. Investment managers and financial institutions benefit from GaR by improving risk management practices and optimising portfolio strategies in different economic scenarios. In addition, government agencies and analysts use GaR to anticipate the impact of economic shocks on public finances.

In this context, it is essential to provide well-calibrated estimates of the quantiles for the future GDP growth distribution.  Standard algorithms, such as the quantile regression (QR) \citep{koenker_regression_1978} or the quantile random forest  (QRF) \citep{meinshausen_quantile_2006}, are commonly used for these purposes, but can produce poorly calibrated results, particularly in the tails of the distribution. The tails, especially the left tail, are of significant interest to policy makers who want to mitigate economic downturns.
%To address these challenges, we adopt Conformal Prediction (CP) \citep{vovk_algorithmic_2023,fontana_conformal_2023}, enhancing the reliability of quantile estimates by providing valid probabilistic guarantees regardless of the underlying distribution. 
To address these challenges, we develop a novel approach that combines Conformal Prediction (CP) \citep{vovk_algorithmic_2023,fontana_conformal_2023} with quantile estimation, providing valid probabilistic guarantees without requiring distributional assumptions. Our method offers three key advantages for practitioners: (1) robust performance under non-standard conditions like fat tails and high dimensionality, (2) automatic adaptation to time-varying uncertainty, and (3) theoretical guarantees on quantile calibration that hold even in finite samples.

Our paper makes three main contributions. First, we develop a novel conformal quantile estimation method with theoretical guarantees. Second, we demonstrate through extensive simulations that our method maintains good calibration even under challenging conditions like fat tails and non-exchangeable time series. Third, we provide empirical evidence that our method improves GaR estimates for US GDP growth, with particular gains in tail risk assessment.

CP methods, originally designed to compute prediction intervals, have not yet been explored for estimating quantiles. In our work, we adapt and extend a conformal algorithm to create a conformalised quantile estimation procedure. Our method, while being a straightforward extension of CP, offers several advantages: it does not rely on stringent assumptions about the data, adapts dynamically to varying levels of uncertainty, and provides theoretical guarantees on the calibration of the estimated quantiles (under certain conditions).
To the best of our knowledge, this is the first study to propose a conformalisation of quantile estimation procedures. %In fact, much of the research in Conformal prediction tends to focus on the production of prediction intervals. 

Our proposal is tested via an extensive simulation study involving time-series data. We explore ill-behaved distributions, characterised by fat tails and high dimensions. The simulation studies encompass scenarios where the innovations lack finite moments, such as Cauchy-distributed errors, and involve a large number of exogenous covariates. In the appendix, we present additional experiments that incorporate fat tails in the error term, modelled by the Student's $t$ distribution with 2 degrees of freedom, and examine the behaviour of the proposed algorithm in situations where the dependent process is highly persistent, explosive, or possesses a unit root. We study the calibration properties of the proposed algorithm in terms of mean absolute error (MAE) of the estimated coverages and the confronting calibration curves. The most striking results appear in the presence of exogenous covariates, which detriment the calibration of fitted quantiles through QR. Interestingly, QRF does compare well with the corresponding conformalised quantiles in terms of empirical coverage at isolated points of the calibration curve. However, in general it does worse in terms of MAE. 

% PARLA DEL FATTO CHE I DATI NON SONO SCAMBIABILI, CHE CIÒ NONOSTATE, SE IL CALIBRATION SET INCREASE LE GARANZIE IID SONO MANTENUTE, CITA OLIVEIRA. NOI COMPLEMENTIAMO QUELLO STUDIO, MOSTRANDO ULTERIORI SETTINGS, DOVE, NONOSTANTE I DATI NON SONO SCAMBIABILI, I RISULTATI SONO BELLI. INOLTRE COMPLEMENTIAMO LA LETTERATURA SU BACKTESTING DEL MARGINAL GAR.
Time series data are inherently non-exchangeable, whereas exchangeability is typically required to ensure that conformal prediction intervals provide valid coverage guarantees. In recent work, \cite{oliveiraetal2024_splitcp} analytically prove that split CP algorithms retain their desirable properties even when applied to non-exchangeable data. Our simulation studies build upon and extend their findings by showing that a modified version of the split CP algorithm for quantile estimation also achieves reliable empirical coverage in the presence of non-exchangeable data (where the direction of non-exchangeability is along the temporal dimension).

The objective of the empirical exercise is to illustrate how our straightforward proposed procedure can enhance the calibration of conditional quantile estimates in real-world case studies.
Hence, we compare our approach to the GaR analysis of  \cite{adrian_vulnerable_2019,adrian_term_2022} and show that conformalisation can provide a better calibrated tool in the hands of practicioners and decision-makers to learn about the future conditional distribution of GDP. We show that conformalised quantiles are better calibrated and approximate the desired coverage both at short (one-quarter ahead) and medium (one-year ahead) horizons. 

Willing to show the appeal of conformalised quantile estimation in a broader class of real-world case studies, we extend the contribution of \cite{adrian_vulnerable_2019} and disaggregate the National Financial Conditions Index (NFCI), a proxy of national financial conditions in the United States (US) updated by the Federal Reserve Bank of Chicago. 
We show that our method captures important nonlinearities in the relationship between financial conditions and growth risks that aggregate indices miss. The enhanced calibration is particularly valuable for policy applications, where reliable tail risk estimates are crucial for decision-making.

In addition to the presented GaR, our conformalised quantile estimation procedure can be applied in any context where calibration of quantile estimates is key, such as financial risk management, which makes extensive use of of Value-at-Risk (VaR) \cite[][Chapter 2]{mcneil_quantitative_2010}. a critical measure used in financial risk management. Accurate VaR calculations are crucial for financial institutions to assess their exposure to risk and determine the necessary capital reserves to cover potential losses. This measure is integral to the Basel III regulatory framework, which requires banks to maintain a certain level of capital reserves proportional to the quantified risk, thus ensuring stability and reducing the likelihood of insolvency in times of financial stress.

There is large and growing empirical literature that uses the GaR framework to understand the downside risks to GDP. In particular,  \cite{adrian_vulnerable_2019} focuses on adverse financial conditions, whereas \citep{adrian_term_2022} study the longer-term effects of adverse financial conditions. We relate our work to \cite{colombeetal2022_MLmacroforecast}, who run a horse-race of machine learning methods in both data-rich and data-poor environments. In the empirical exercise, they also propose conditional quantile estimation, but their estimator does not posses any property that might guarantee satisfying calibration. In addition, our results speak to \cite{brownlees2021backtesting}, who perform backtesting of GaR and show that models with conditional heteroscedasticity (GARCH) produce forecasts that are more accurate compared to quantile regression. Our results extend the literature on backtesting of marginal GaR, providing a better calibrated quantile estimation, while we do not focus on accuracy of our estimates. The comparison between conformalised quantile regression and GARCH models, or even a conformalisation of conditional heteroschedasticity models, is left as an interesting venue for future research. 

The paper is structured as follows. Section \ref{sec:methodology} presents the methodology, introducing the Conformalised Quantile Regression algorithm and its modification and adaptation for the estimation of quantiles. Section \ref{sec:simulation} is dedicated to the simulation study with generated data. Section \ref{sec:realcase} is reserved for empirical illustration, estimating future GDP growth in the United States.

\section{Methodology}
\label{sec:methodology}
CP is a powerful, model-agnostic framework for constructing prediction intervals that provide a guaranteed coverage probability, regardless of the underlying data distribution. Unlike traditional approaches that rely on specific distributional assumptions, CP leverages the concept of conformity scores to measure the ``strangeness'' of new observations relative to a given dataset. Given a dataset $\mathcal{D} = \{(X_t, Y_t)\}_{t=1}^n$, a conformity score function $E_t:= E(X_t, Y_t)$ is defined to quantify how unusual is each observation $(X_t, Y_t)$.

In this paper, we focus on a specific CP method using a split CP approach, where the dataset is partitioned into two disjoint subsets: one for training the predictive model and another for calibration. The conformity scores are computed on the calibration set, and an empirical quantile of these scores is used to construct prediction intervals. We propose a novel modification to extend the applicability of this method from constructing prediction intervals to directly estimating quantiles.
To the best of our knowledge, this is the first work in the literature that uses a CP framework to estimate quantiles directly rather than to construct prediction intervals. This enhancement broadens the utility of CP, making it suitable for applications where understanding the entire distribution is crucial.

\subsection{Conformalised Quantile Regression and Quantile Estimation}
\subsubsection{Conformalised Quantile Regression}
Conformalised Quantile Regression (CQR) is an algorithm introduced in \cite{romano_conformalised_2019}. 
It is an advanced statistical technique that combines CP methods with quantile regression to provide robust prediction intervals with valid coverage guarantees in finite samples.
The algorithm begins with a split of the dataset in two disjoint subsets, \( \mathcal{T}_1 \) and \( \mathcal{T}_2 \).
Given the conditional quantile function $Q(\tau, x) := \inf\{y\in \mathbb{R}| \mathbb{P}(Y\leq y| X=x)\geq \tau\}$, with $\tau\in [0,1]$, subset \( \mathcal{T}_1 \) is used to fit two conditional quantile functions \( \hat{Q}(\alpha/2, x) \) and \( \hat{Q}(1 - \alpha/2, x) \), where \( \alpha \) is the miscoverage rate (i.e. prediction intervals of level \(1 -  \alpha \) will be computed), and hats indicate fitted values. The algorithm is flexible regarding how these  fitted values are computed. In the following, we focus on two popular choices, namely quantile regression and quantile random forest. 

As in \cite{romano_conformalised_2019}, the following guarantees about coverage  hold:
\begin{theorem}[\cite{romano_conformalised_2019}]\label{thm:2.1}
If $(X_t, Y_t)$, $t = 1, \ldots, n+1$ are exchangeable, then the prediction interval $C(X_{n+1})$ constructed by the split CQR algorithm satisfies:
\begin{equation}
\mathbb{P}\{Y_{n+1} \in C(X_{n+1})\} \geq 1 - \alpha. \label{eq:thm21_prop1}
\end{equation}

Moreover, if the conformity scores $E_t$ are almost surely distinct, then the prediction interval is nearly perfectly calibrated:
\begin{equation}
\mathbb{P}\{Y_{n+1} \in C(X_{n+1})\} \leq 1 - \alpha + \frac{1}{|\mathcal{T}_2| + 1}. \label{eq:thm21_prop2}
\end{equation}
\end{theorem}

\subsubsection{CQR for Quantile Estimation}
The objective of our paper is to propose a method that provides estimated quantiles, rather than prediction intervals for forecast $Y_{n+1}$. Some modifications to the algorithm of \cite{romano_conformalised_2019} are needed to produce quantile estimates using CP. 
We consider two approaches. % we can take to achieve our goal.
In the first method, we start by estimating only the upper bound of the prediction interval, that is the quantile function of level \(1 -  \alpha \): \( \hat{Q}(1 - \alpha, x) \). It can be interpreted that the lower quantile function is of level 0, resulting in \( \hat{Q}(0, x) = -\infty \).
Therefore, the calculation of conformity scores is simply
\[
E_t = \max \left\{ -\infty, \, Y_t - \hat{Q}(1 - \alpha, X_t) \right\} = Y_t - \hat{Q}(1 - \alpha, X_t).
\]
The resulting estimate of the quantile of level \(1 -  \alpha \) then follows from the upper bound of the prediction interval
\[
C(x) = \left( -\infty, \, \hat{Q}(1 - \alpha, x) + Q_E(1 - \alpha) \right].
\]
In our quantile setting, equation \eqref{eq:thm21_prop1} reads
\[
\mathbb{P}\{Y_{n+1} \leq \hat{Q}(1 - \alpha, x) + Q_E(1 - \alpha)\} \geq 1 - \alpha,
\]
which is, in fact, the definition of a quantile, available in any basic probability theory textbook. % In layman's terms, it means that estimate will exceed the next observation at least \((1 -  \alpha)\% \)  of the times.
However, in GaR analysis, policymakers are interested in a different inequality, that is one that allows for controlling the frequency of observations falling below a specified threshold (as opposed to above the threshold).\footnote{This perspective on the algorithm is aligned with the definition of VaR at the $\alpha$ confidence level in financial risk management, $$VaR_{\alpha}(X) := \underset{u\in \mathbb{R}}{\inf}\{u\in \mathbb{R}:\mathbb{P}(X \leq u)\leq\alpha\},$$ i.e., the largest loss 
$u$ such that the probability of experiencing a loss larger (more negative) than $u$ is at most $\alpha$.} To address this need, we introduce a second procedure for quantile estimation.
We now produce only the lower quantile function of level \(\alpha \):  \( \hat{Q}(\alpha, x) \). The upper quantile function can be seen as \( \hat{Q}(1, x) = \infty \):
\[
E_t = \max \left\{ \hat{Q}(\alpha, X_t) - Y_t, \, -\infty \right\} = \hat{Q}(\alpha, X_t) - Y_t,
\]
and consequently
\[
C(x) = \left[ \hat{Q}(\alpha, x) - Q_E(1 - \alpha) , \, +\infty \right).
\]
And now, according to (\ref{eq:thm21_prop1}), we conclude
\[
\mathbb{P}\{Y_{n+1} \geq \hat{Q}(\alpha, x) - Q_E(1 - \alpha)\} \geq 1 - \alpha,
\]
correspondingly, 
\begin{equation}
\mathbb{P}\{Y_{n+1} \leq \hat{Q}(\alpha, x) - Q_E(1 - \alpha)\} \leq \alpha,
\label{eq:main_eq}    
\end{equation}
where, we assumed continuity of \(\hat{Q}(\alpha, x)\). For  fixed \(X= x\), \(\hat{Q}(\alpha, x)\) corresponds to the inverse conditional cdf of \(Y\) given \(X=x\), which is guaranteed to be continuous when the model includes a continuous error term, as is the case in all simulations conducted in Section \ref{sec:simulation}. For fixed \(\alpha\), the continuity of \(\hat{Q}(\alpha, x)\) with respect to \(X\) depends on the smoothness of the relationship between \(Y\) and \(X\). In all our simulations, and in most practical scenarios, this condition is satisfied. Due to its useful coverage guarantee, this second procedure is employed throughout this paper.

Inequality \eqref{eq:main_eq} is exactly what is needed to control the frequency of extreme adverse outcomes in GDP growth in the future. This formulation enables, for example, central banks to establish a predefined threshold for GDP growth (e.g., 2\%) and to quantify the maximum probability of experiencing an adverse or worse outcome. In contrast, the inequality can also be used to identify the specific GDP growth value associated with a given risk level \(\alpha\). This dual application is crucial for policy formulation, as it provides a robust framework for managing economic risks by aligning GDP growth projections with acceptable risk thresholds.
 
Lastly, it is important to highlight the main limitation of this property: the exchangeability assumption. The assumption is naturally violated in time series data. In the absence of exchangeability, the result stated in (\ref{eq:main_eq}) does not hold. 

Since exchangeability is a crucial assumption in many CP algorithms, new techniques and methods are being developed to address this limitation. However, these techniques won't be explored in this paper; instead, we focus on assessing whether this property holds, even approximately, in both simulated and real data when data are not exchangeable.

\section{Simulation Study}
\label{sec:simulation}
In order to stress and test the robustness of standard methods for quantile estimation, as well as to assess the effectiveness of conformal methods, we conduct a comprehensive simulation study, using synthetic data. Since the main focus of our work is the economic and financial sector, the simulations  involve time series with autoregressive components, with fat-tailed distributions and/or high dimensionality. 
Simulations are all built in a similar fashion. A time series \(\left\{(X_i, Y_i)\right\}_{i=1}^{n+100}\) is generated for \( n \in \left\{98, 198, 998\right\}\). The first \(n\) observations are  used to train (and calibrate) the models, while the last 100 observations are be used to evaluate the models' performance.
Performances are then averaged over 100 iterations, which corresponds to generating 100 different time series as just described.
In each iteration, the models are trained to predict the entire distribution of the target variable \({Y}_{i}\) through the prediction of 20 quantile levels \({L}_i \in \left\{0.01, 0.05, 0.10, 0,15, ..., 0.95\right\}\). For each quantile level in each iteration the unconditional empirical coverage \(\hat{P}(\hat{Q}_i)\) is calculated, which is simply the proportion of test data that falls below the empirical quantile value \(\hat{Q}_i\) estimated by our model. In a perfectly calibrated model, the coverages are equal to their associated quantile level: the estimated quantile of level 0.9 should be greater than 90\% of test data.

Four different models are implemented and tested: Quantile Regression (QR), Quantile Random Forest (QRF) and their CQR versions. 
QR and QRF offer complementary strengths and differ in their underlying assumptions, interpretability, and flexibility.
QR is a parametric technique that allows for the estimation of specific quantiles of the dependent variable, given a set of predictors.
It assumes a linear relationship between the predictors and the quantiles of the response. Its interpretability makes it a popular tool in the economic field.
On the other hand, QRF extends the traditional Random Forest algorithm to provide non-parametric estimates of conditional quantiles. Unlike QR, QRF does not assume a linear relationship between predictors and the response, allowing it to capture complex, non-linear interactions within the data. Given these differences, employing both Quantile Regression and Quantile Random Forest allows for a comprehensive comparison of linear versus non-linear effects, interpretability versus flexibility, and parametric versus non-parametric methods. QR and QRF then are also used to fit a conditional quantile function \(\hat{Q}(\alpha, x)\) to perform the CQR algorithm, obtaining the CQR QR and the CQR QRF models.

Performances are evaluated estimating the mean absolute error between the estimated coverages and the quantile level:
\[
\mathrm{MAE} := \frac{1}{2000} \sum_{i=1}^{20} \sum_{j=1}^{100} \left| \hat{P}(\hat{Q}_{ij}) - {L}_{i} \right|.
\tag{4}\label{eq:MAE}
\]
The quantity is averaged across all the iterations and quantiles. Perfectly calibrated models have \(\mathrm{MAE} = 0\) and a low value of \(\mathrm{MAE}\) is associated with a well calibrated model. However, \(\mathrm{MAE}\) alone is not sufficient to fully evaluate a model's performance, as it does not distinguish whether the error is due to \(\hat{P}(\hat{Q}_i) > {L}_i\) or to a conservative coverage \(\hat{P}(\hat{Q}_i) < {L}_i\). To address this limitation, we construct a 95\% binomial proportion confidence interval, calculated using the Wilson score, centred around the empirical coverage \(\hat{P}(\hat{Q}_i)\). This procedure allows us to construct a confidence interval for the real quantile level \({L}_i\). In each of the simulation sections below, a table summarises these results. These tables provide the percentage of quantile levels \({L}_i\) that fall within, below, or above the confidence interval constructed around their empirical coverage. 
"Within CI" refers to the percentage of quantile levels \({L}_i\) that fall within this confidence interval, and we consider these estimates as perfectly calibrated. "Below CI" and "Above CI" represent the percentages of quantile levels \( L_i \) that fall below or above, respectively, the constructed confidence interval. 
If (\ref{eq:main_eq}) holds, we expect for CQR algorithms a small \% in "Below CI", lower when compared to their respective standard algorithms.
Additionally, we include calibration curves to provide an immediate and insightful visual representation.
Two different data generating processes are explored in the following subsections, while a third is available in Appendix \ref{subsec:unitroot}.

\subsection{AR(2) with Cauchy distributed errors}
\label{subsec:cauchy}

As a first simulation, the following Data Generating Process (DGP) is implemented:
\[
Y_t = \phi_1 Y_{t-1} + \phi_2 Y_{t-2} + \epsilon_t
\]
where
\[
 \phi_1 = 0.5, \quad \phi_2 = -0.2, \quad \epsilon_t \sim \text{Cauchy}(0, 1).
\]

The introduction of a Cauchy error term significantly impacts the characteristics of an AR(2) model. Specifically, it renders the moments (such as mean, variance, skewness) undefined, thereby posing substantial challenges to standard prediction models. The heavy-tailed nature of the Cauchy distribution, characterized by frequent outliers, makes it particularly suitable for modelling financial data. This is especially relevant in the field of risk management (GaR, for example), where accurately capturing the probability of extreme losses induced by outliers is crucial.

\begin{figure}[H]
    \centering
    \subfloat[\label{fig:n98_Cauchy}]{
        \includegraphics[width=0.29\textwidth]{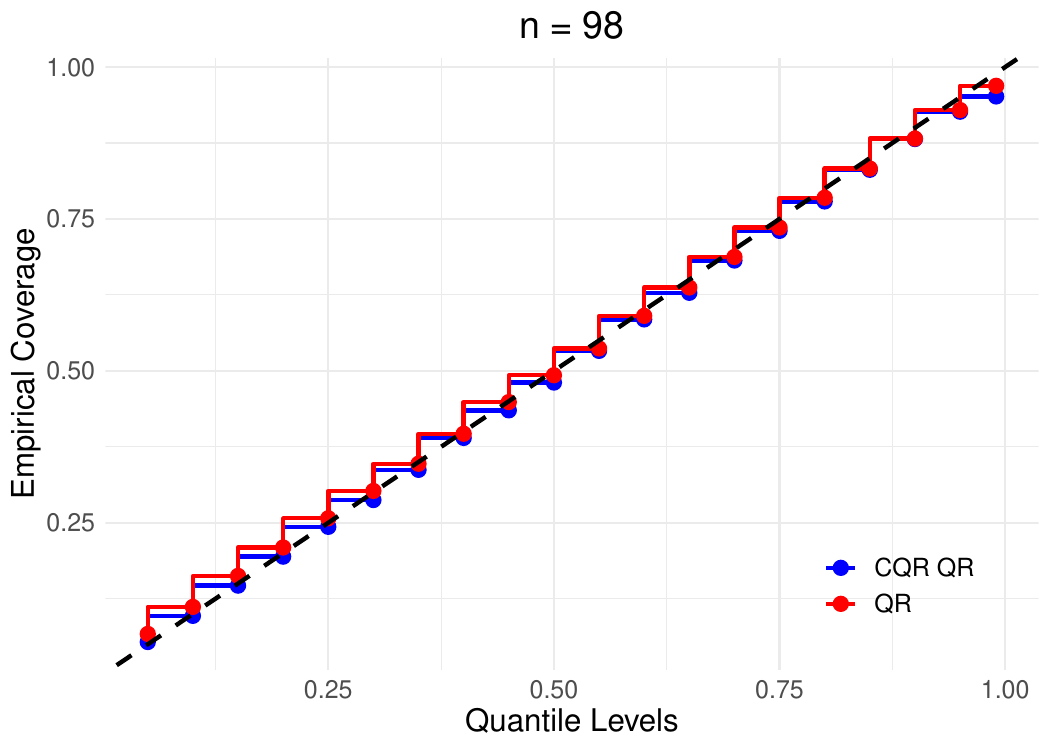}
    }
    \hspace{1em}
    \subfloat[\label{fig:n198_Cauchy}]{
        \includegraphics[width=0.29\textwidth]{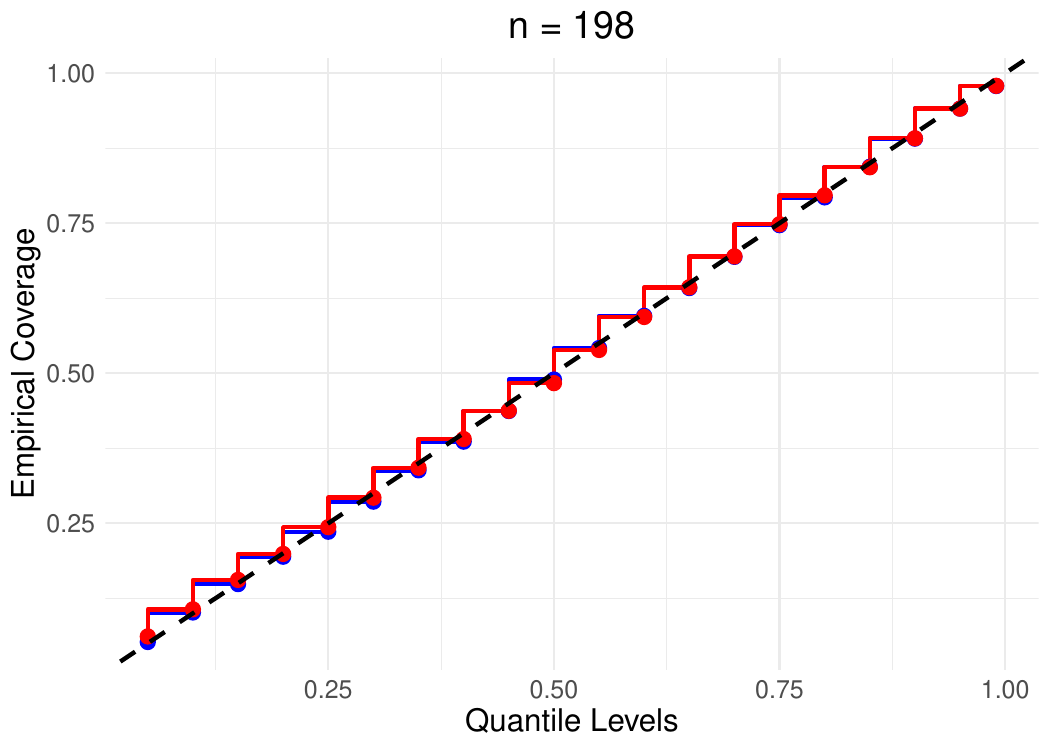}
    }
    \hspace{1em}
    \subfloat[\label{fig:n998_Cauchy}]{
        \includegraphics[width=0.29\textwidth]{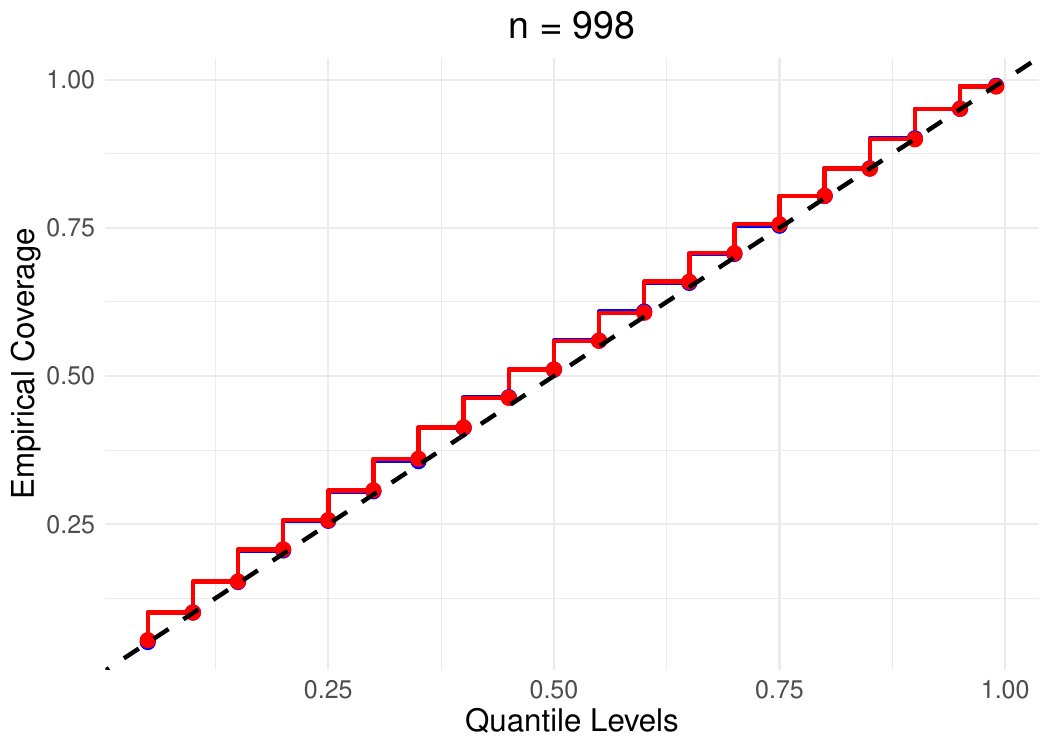}
    }
    \captionsetup{font=footnotesize}
    \caption[]{Calibration curves for QR and CQR QR for the 3 values of \(n\).}
    \label{fig:figure1}
\end{figure}

\begin{figure}[H]
    \centering
    \subfloat[\label{fig:n98_Cauchy.}]{
        \includegraphics[width=0.29\textwidth]{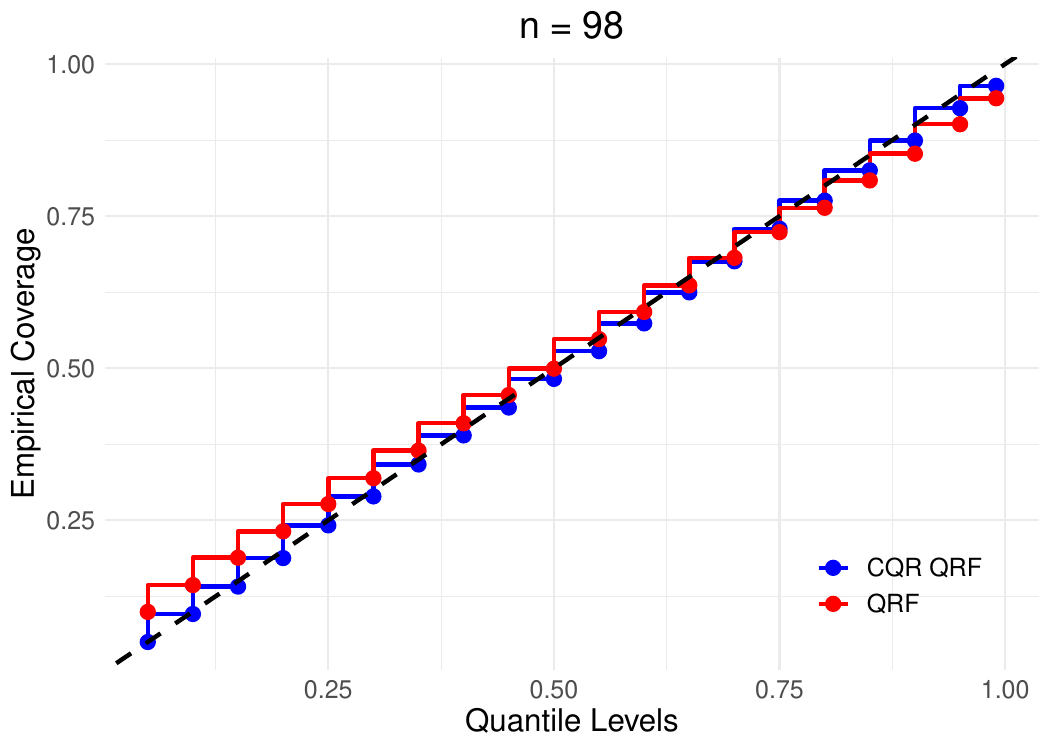}
    }
    \hspace{1em}
    \subfloat[\label{fig:n198_Cauchy.}]{
        \includegraphics[width=0.29\textwidth]{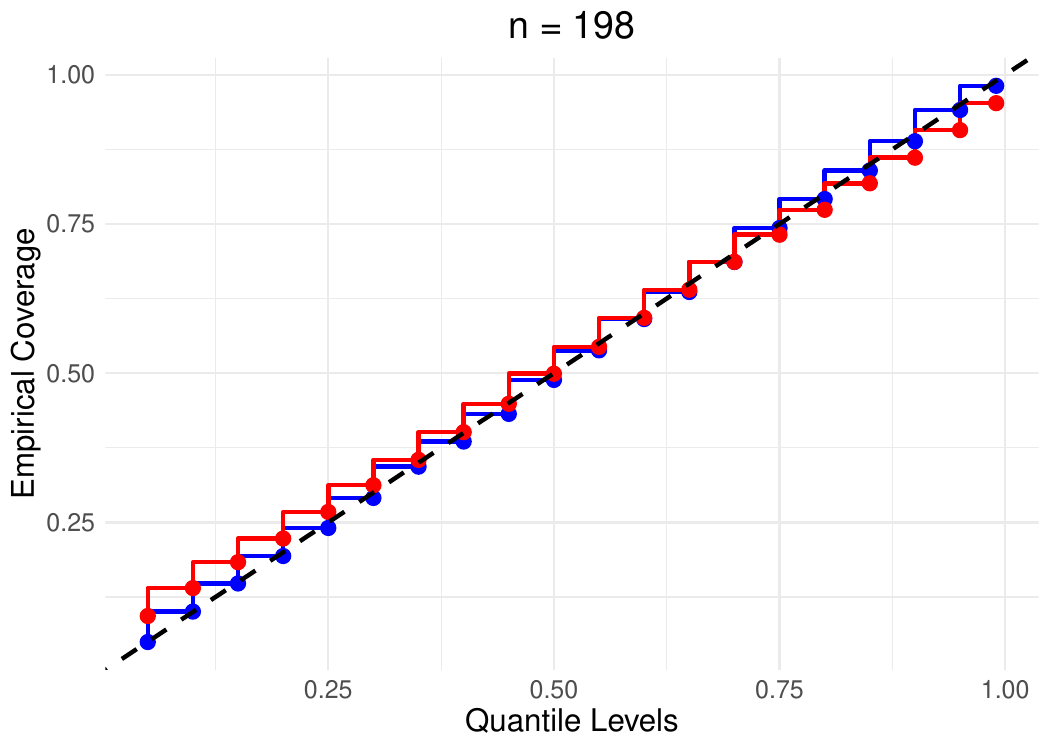}
    }
    \hspace{1em}
    \subfloat[\label{fig:n998_Cauchy.}]{
        \includegraphics[width=0.29\textwidth]{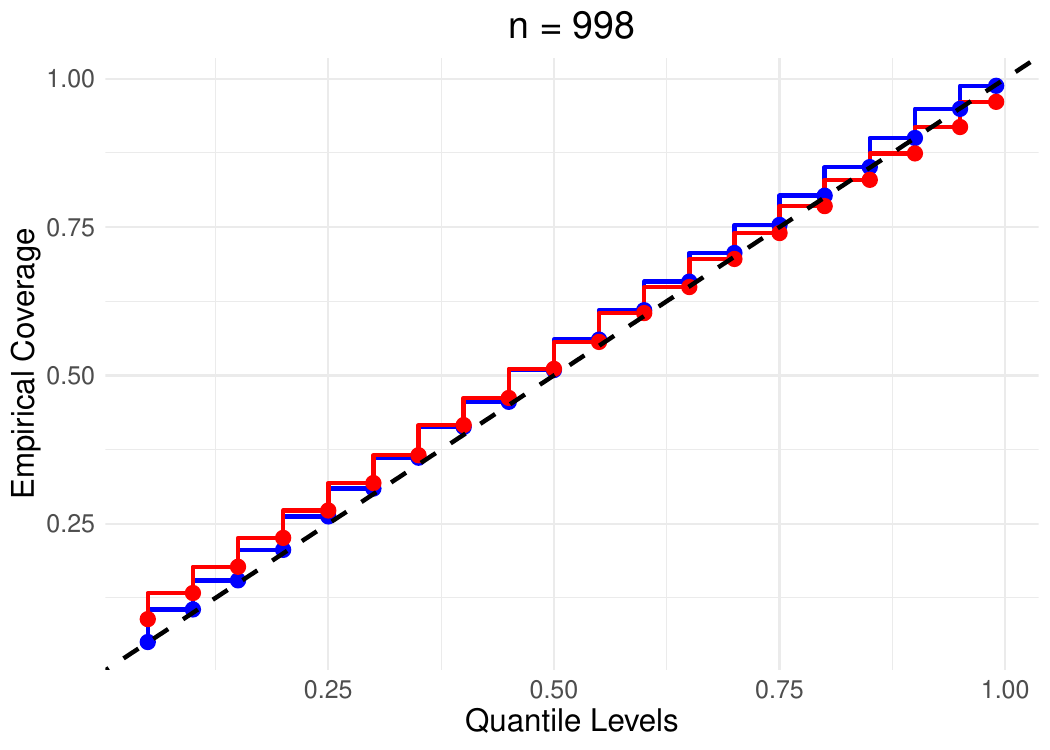}
    }
    \captionsetup{font=footnotesize}
    \caption[]{Calibration curves for QRF and CQR QRF for the 3 values of \(n\).}
    \label{fig:figure2}
\end{figure}

The results presented in Table \ref{table:table1} show that QR reaches a satisfactory calibration even without conformalisation, while the latter significantly improves the calibration for QRF. More importantly, the table shows that property (\ref{eq:main_eq}) is valid across almost all quantile levels for CQR algorithms. This was not guaranteed because the exchangeability hypothesis is not met in this model. Table \ref{table:table2} further confirms an improvement in performance for CQR QRF, showing a decrease in \(\mathrm{MAE}\), while the performances of QR and CQR QR are overall similar.
The calibration curves in Figures \ref{fig:figure1} and \ref{fig:figure2} highlight which quantile levels \({L}_i\) are better calibrated. As previously noted, QR estimations are accurate even without conformalisation. In contrast, the QRF algorithm yields poorly calibrated estimates, particularly at extreme quantile levels, such as 0.05 or 0.95. Since GaR specifically targets the lowest quantile of the GDP growth distribution, the use of conformalisation techniques becomes even more critical.

\begin{table}[H]
    \caption*{\textbf{}}
    \centering 
    \begin{tabular}{|p{6em} c c c |}
    \hline
    \rowcolor{lavender}
     & \textbf{Within CI} & \textbf{Below CI} & \textbf{Above CI}  \\
    \hline \hline
    \textbf{QR} & 58.33\% & 16.67\% & 25.00\%  \\
    \textbf{QRF} & 25.00\% & 38.34\% & 36.66\%  \\
    \textbf{CQR QR} & 55.00\% & 5.00\% & 40.00\%  \\
    \textbf{CQR QRF} & 40.00\% & 10.00\% & 50.00\%  \\
    \hline
    \end{tabular}
    \\[10pt]
    \captionsetup{font=footnotesize}
    \caption{ The table contains the percentage of quantile levels \({L}_i\) which falls within, above or below the 95\% binomial proportion confidence interval calculated using Wilson Score, built around the empirical coverage \(\hat{P}(\hat{Q}_i)\). Results are calculated over the 3 values of \(n\).}
    \label{table:table1}
\end{table}

\begin{table}[H]
    \caption*{}
    \centering 
    \begin{tabular}{|p{7em} c c c|}
    \hline
    \rowcolor{lavender}
      \textbf{MAE} & \textbf{98} & \textbf{198} & \textbf{998}  \\
    \hline \hline
    \textbf{QR} & 0.011 & 0.008 & 0.006 \\
    \textbf{QRF} & 0.026 & 0.021 & 0.018 \\
    \textbf{CQR QR} & 0.015 & 0.008 & 0.005 \\
    \textbf{CQR QRF} & 0.017 & 0.009 & 0.006 \\
    \hline
    \end{tabular}
    \\[10pt]
    \captionsetup{font=footnotesize}
    \caption{MAE (\ref{eq:MAE}) for the 4 models and the 3 values of \(n\)}
    \label{table:table2}
\end{table}

\subsection{AR(2) with Exogenous variables}
\label{subsec:exogenous}
As a second simulation, the following DGP is implemented:
\[Y_t = \phi_1 Y_{t-1} + \phi_2 Y_{t-2} + \bm{\beta}^\top \mathbf{X}_t + \epsilon_t\]
where
\[\phi_1 = 0.5, \quad \phi_2 = -0.2, \quad \epsilon_t \sim t_2,\]
\[\beta_i \sim \text{\textit{U}}(0, 1) \quad \text{for each} \quad i = 1, \ldots, p, \quad p/n \in \left\{0.1, 0.2, 0.3, 0.4\right\}\]

\[
\mathbf{X}_t \sim \mathcal{N}_p(\textbf{m}, \bm{\Sigma})
\]
\vspace{-4pt}
\[
\textbf{m} \sim \mathcal{N}_p(\textbf{0}, \textbf{I}), \quad 
\bm{\Sigma} =
  \begin{bmatrix}
    \sigma_{1} & & \\
    & \ddots & \\
    & & \sigma_{p}
  \end{bmatrix} , \quad \sigma_i \sim \text{\textit{U}}(0, 10) 
\]

In this second simulation, we again use a fat-tailed error in the DGP, this time a t-Student distribution with two degrees of freedom. This error has heavy tails similar to the Cauchy distribution, but unlike the Cauchy, it has a finite mean, although its variance remains undefined. We also introduce a term related to exogenous variables to test the models' calibration when the number of covariates is high. In particular, we train models for 4 different ratios of the number of covariates and observations \(p/n: 0.1, 0.2, 0.3, 0.4\).
Given the complexity of 21st-century economies, identifying the drivers of growth, even retrospectively, is a challenging task that often results in datasets containing a vast number of macroeconomic and financial variables. Although efforts have been made to consolidate many of these variables into a single index, such as the National Financial Conditions Index (NFCI), the ability to manage a large number of covariates is a crucial feature for a model designed for GaR analysis.\footnote{The NFCI is maintained by the Federal Reserve Bank of Chicago and is available at: \url{https://www.chicagofed.org/research/data/nfci/current-data}.}

\begin{figure}[H]
    \centering
    \includegraphics[width=0.6\textwidth]{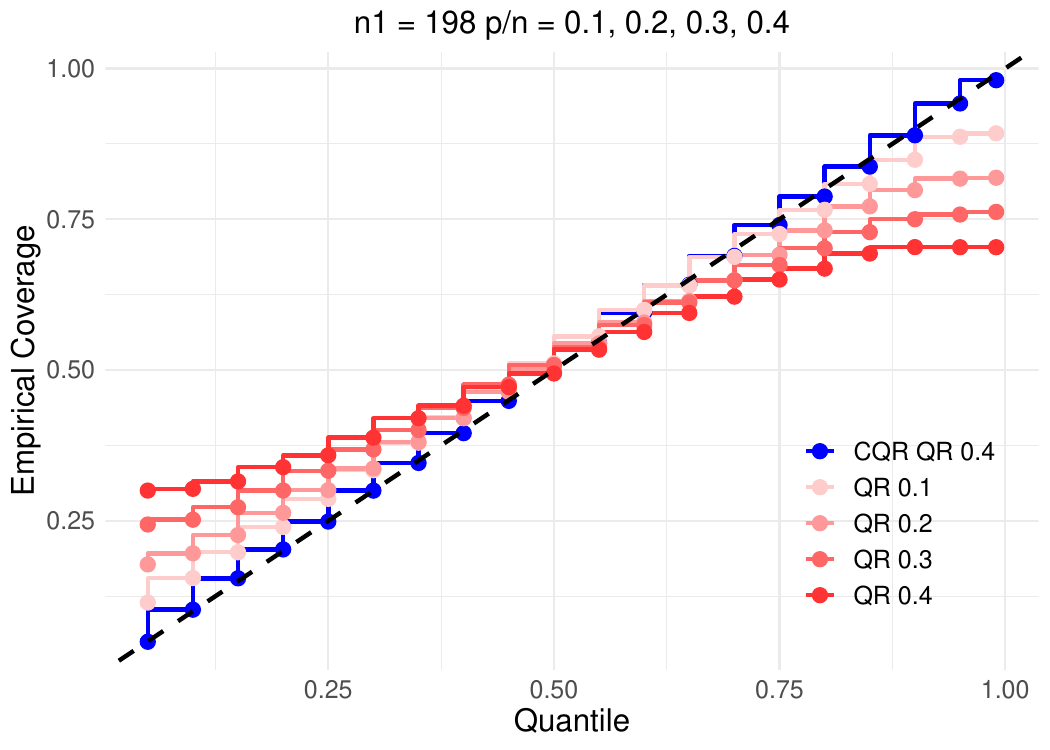}
    \captionsetup{font=footnotesize}
    \caption[]{This plot is intended to provide a quick and intuitive visualization of the calibration of QR models when the ratio \(p/n\) increases, compared to a single CQR QR model  with \(p/n = 0.4\), that is very well calibrated. The individual plots can be seen below in Figure \ref{fig:figure4} and in Appendix \ref{appendix:A.1}.}
    \label{fig:figure3}
\end{figure}

\begin{figure}[H]
    \centering
    \subfloat[\label{}]{
        \includegraphics[width=0.4\textwidth]{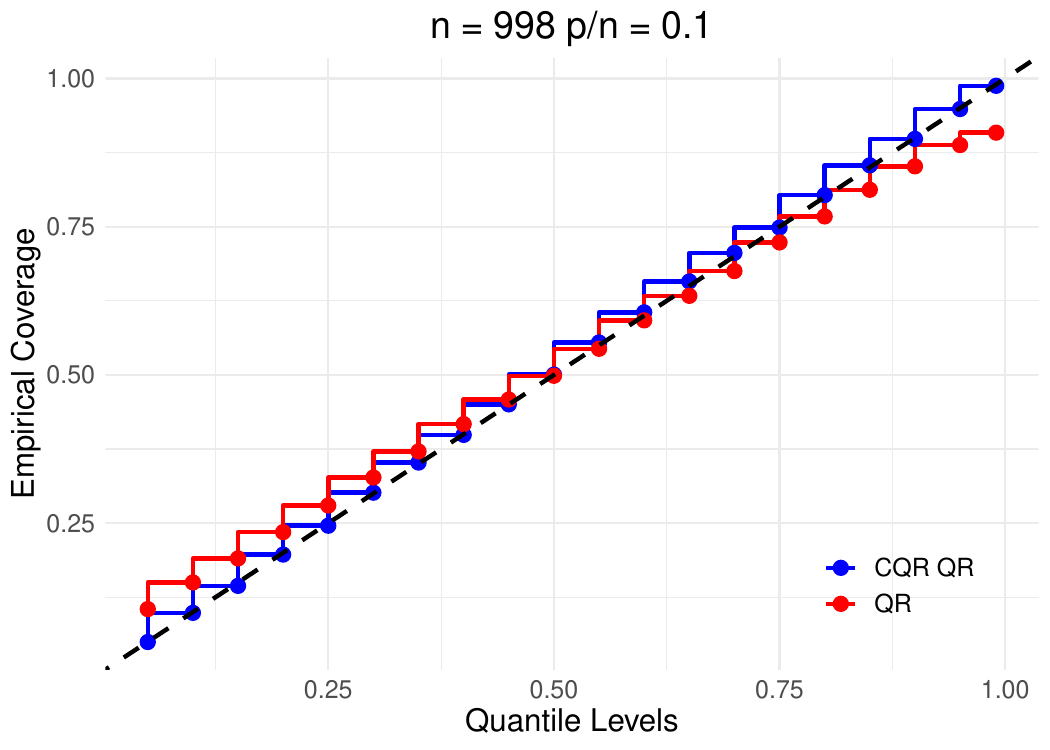}
    }
    \hspace{1em}
    \subfloat[\label{}]{
        \includegraphics[width=0.4\textwidth]{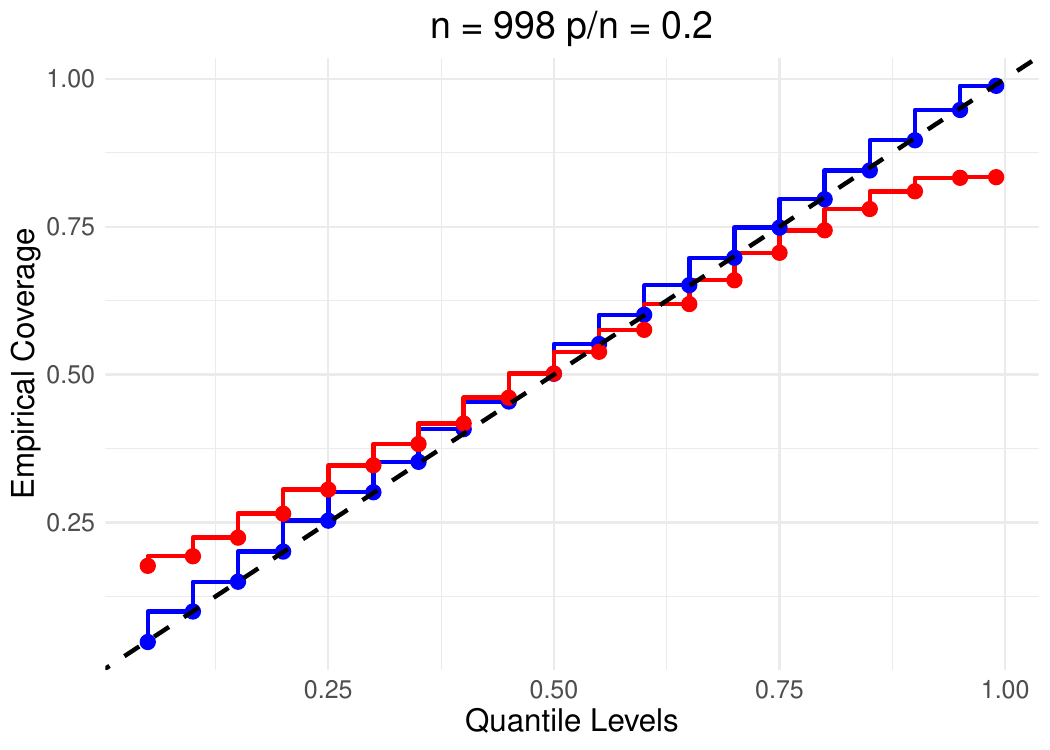}
    }
    \vspace{1em}
    \hspace{1em}
    \subfloat[\label{}]{
        \includegraphics[width=0.4\textwidth]{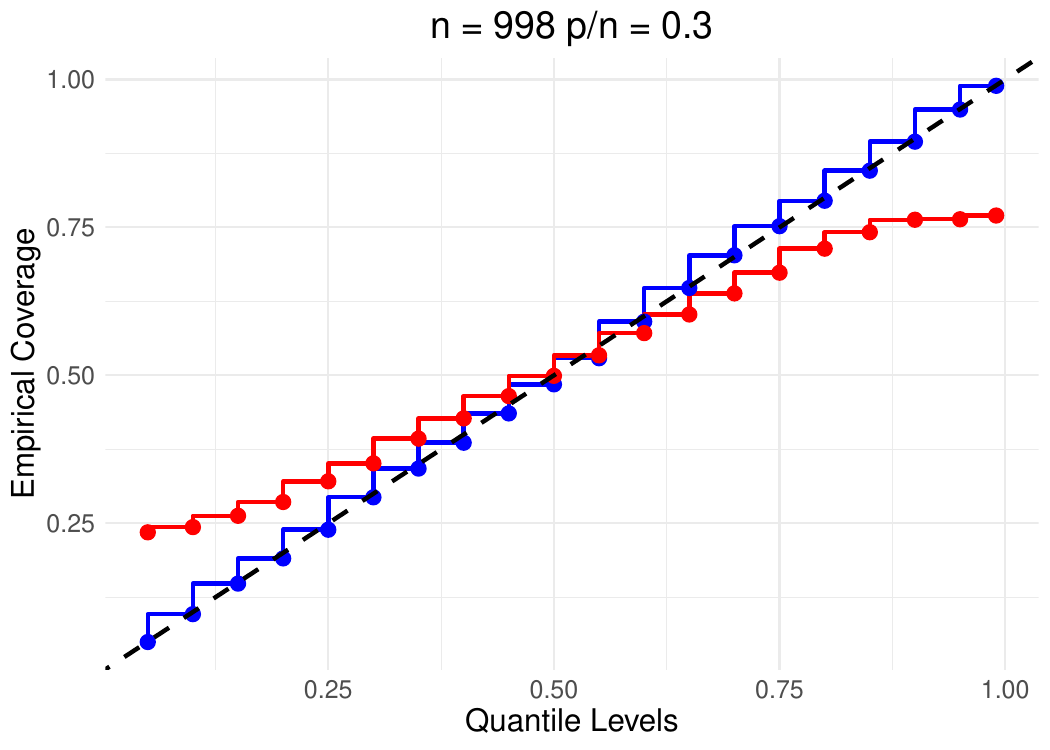}
    }
    \hspace{1em}
    \subfloat[\label{}]{
        \includegraphics[width=0.4\textwidth]{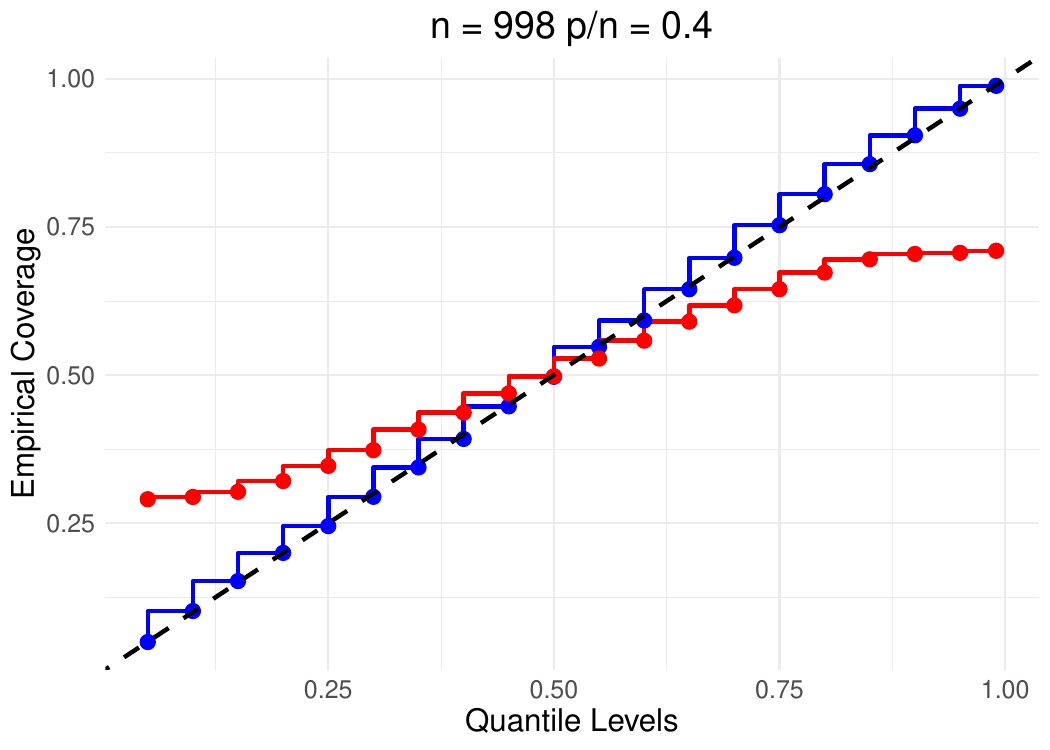}
    }
    \captionsetup{font=footnotesize}
    \caption[]{Calibration curves for QR and CQR QR for \( n = 998 \) and \( p/n \) values 0.1, 0.2, 0.3, and 0.4. Plots are similar for the other two values on \(n\) and can be seen in Appendix \ref{appendix:A.1}}.
    \label{fig:figure4}
\end{figure}

In this simulation, even more than in the first, it is clear that conformalisation significantly enhances model calibration. From the MAE in Table \ref{table:table4} and the calibration curves in Figure \ref{fig:figure3} and \ref{fig:figure4}, we note that QR provides very poor quantile estimations, especially for extreme quantile levels, and calibration worsens as the ratio \(p/n\) increases.
QRF estimations are more precise than QR estimations, but they nevertheless benefit from conformalisation.
CQR methods are also better calibrated for an increasing value of \(n\), while improvement for QR and QRF is either less evident or entirely questionable.
The guarantee (\ref{eq:main_eq}) holds once again, and even more convincingly, with all the positive consequences that have already been discussed (Table \ref{table:table3}).

\begin{figure}[H]
    \centering
    \subfloat[\label{}]{
        \includegraphics[width=0.4\textwidth]{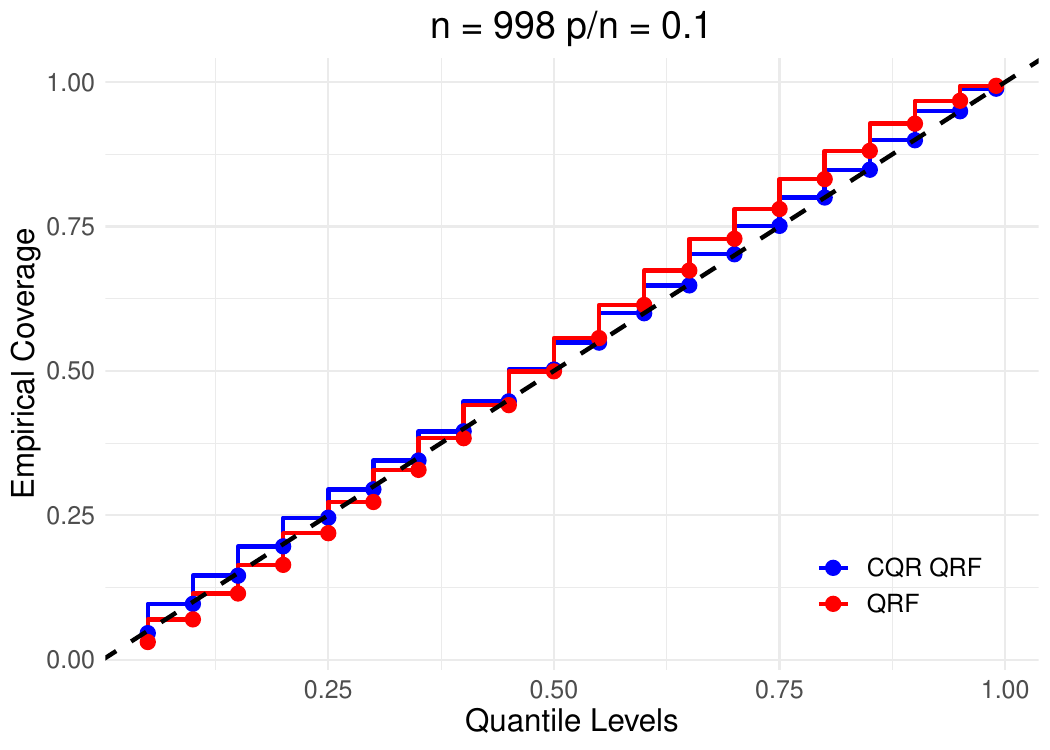}
    }
    \hspace{1em}
    \subfloat[\label{}]{
        \includegraphics[width=0.4\textwidth]{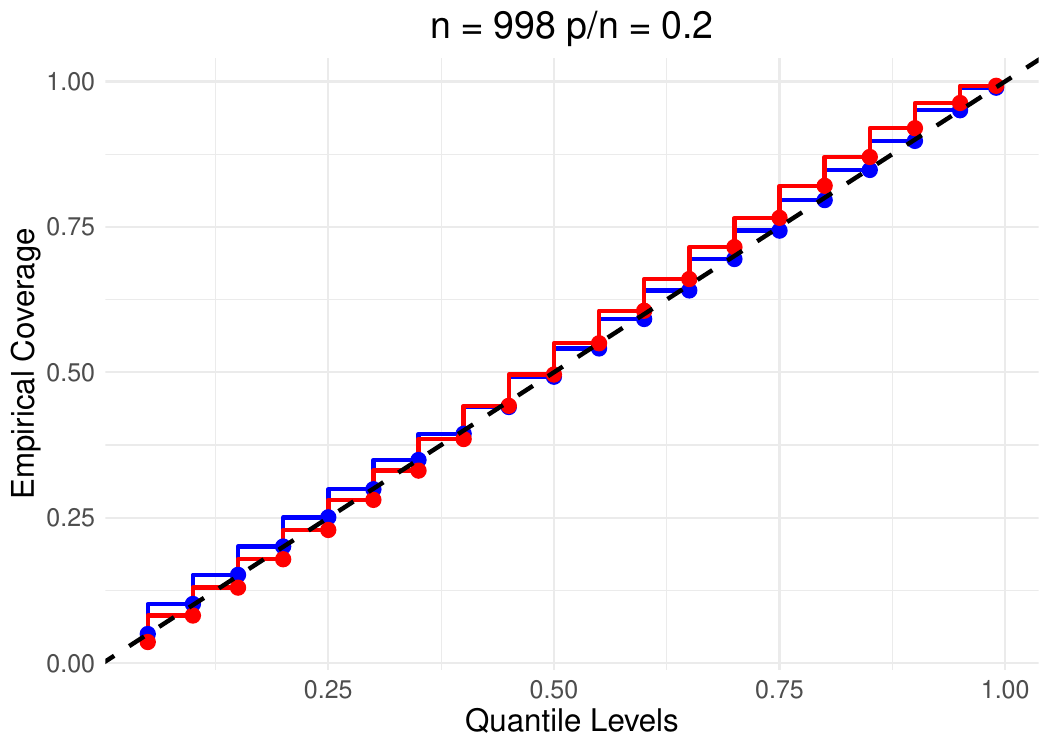}
    }
    \vspace{1em}
    \hspace{1em}
    \subfloat[\label{}]{
        \includegraphics[width=0.4\textwidth]{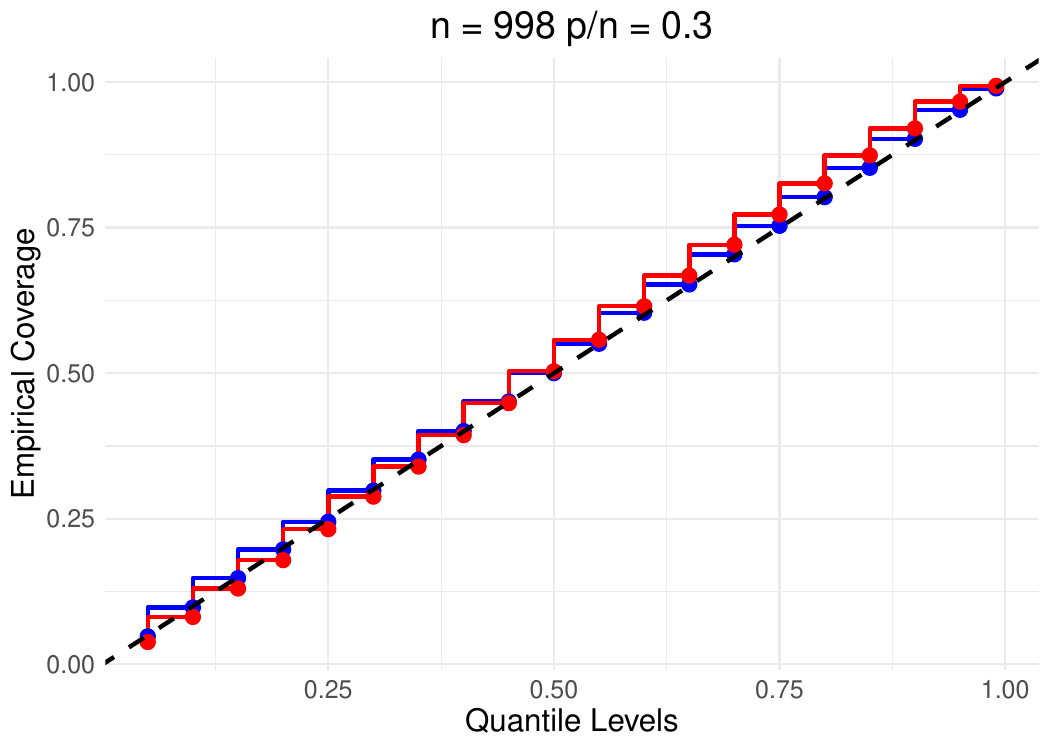}
    }
    \hspace{1em}
    \subfloat[\label{}]{
        \includegraphics[width=0.4\textwidth]{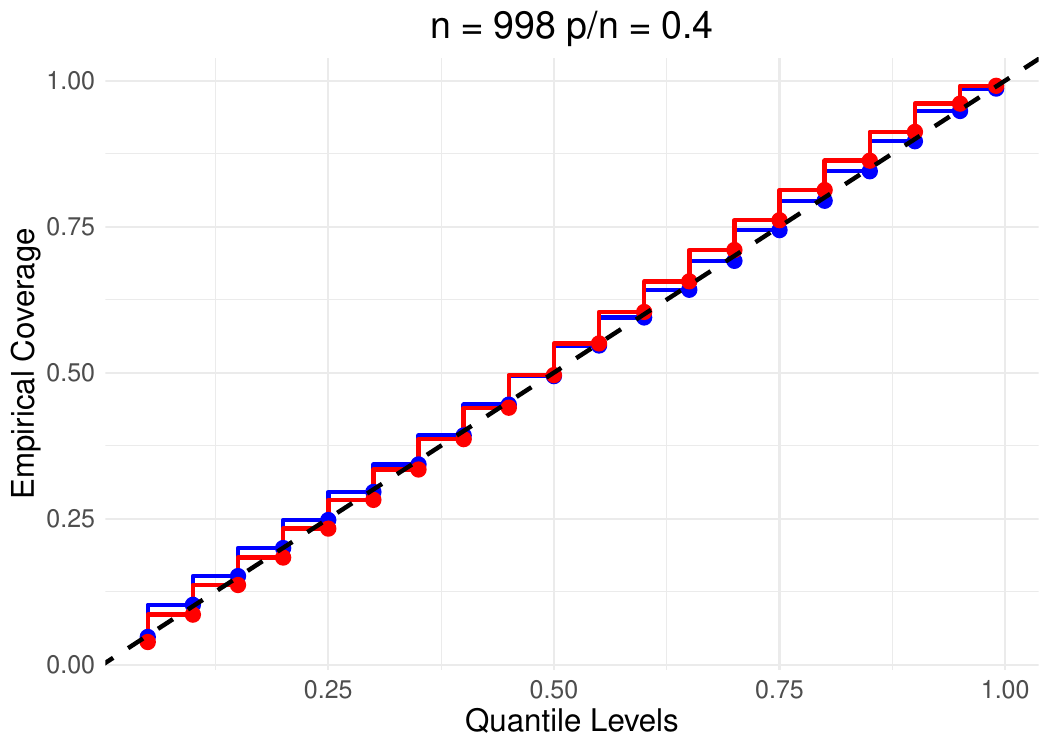}
    }
    \captionsetup{font=footnotesize}
    \caption[]{ Calibration curves for QRF and CQR QRF for \( n = 998 \) and \( p/n \) values 0.1, 0.2, 0.3, and 0.4. Plots are similar for the other two values on \(n\) and can be seen in Appendix \ref{appendix:A.1}}.
    \label{fig:figure5}
\end{figure}

\begin{table}[H]
    \caption*{\textbf{}}
    \centering 
    \begin{tabular}{|p{6em} c c c |}
    \hline
    \rowcolor{lavender}
     & \textbf{Within CI} & \textbf{Below CI} & \textbf{Above CI}  \\
    \hline \hline
    \textbf{QR} & 7.92\% & 43.33\% & 48.75\%  \\
    \textbf{QRF} & 18.75\% & 41.25\% & 40.00\%  \\
    \textbf{CQR QR} & 59.16\% & 3.75\% & 37.08\%  \\
    \textbf{CQR QRF} & 70.00\% & 0.83\% & 29.17\%  \\
    \hline
    \end{tabular}
    \\[10pt]
    \captionsetup{font=footnotesize}
    \caption{The table contains the percentage of quantile levels \({L}_i\) which falls within, below or above the 95\% binomial proportion confidence interval calculated using Wilson Score, built around the empirical coverage \(\hat{P}(\hat{Q}_i)\). Results are calculated over the 3 values of \(n\) and the 4 values of \(p/n\).}
    \label{table:table3}
\end{table}

\begin{table}[H]
    \caption*{}
    \centering 
    \begin{tabular}{|p{7em} c c c|}
    \hline
    \rowcolor{lavender}
      \textbf{MAE} & \textbf{98} & \textbf{198} & \textbf{998}  \\
    \hline \hline
    \textbf{QR} &  0.08 & 0.077  & 0.073  \\
    \textbf{QRF} &  0.019 &  0.02 & 0.015  \\
    \textbf{CQR QR} &  0.012 &  0.007 & 0.004  \\
    \textbf{CQR QRF} &  0.013 &  0.005 & 0.003  \\
    \hline
    \end{tabular}
    \\[10pt]
    \captionsetup{font=footnotesize}
    \caption{MAE (\ref{eq:MAE}) for the 4 models and the 3 values of \(n\), averaged on the 4 values of \(p/n\).}
    \label{table:table4}
\end{table}

It is interesting to note that due to the design of CQR algorithm, only half of the training set is actually used to train a quantile prediction model, while the other half is reserved for calibration. Thus, in practice, CQR QR and CQR QRF models are trained on a effective ratio up to 0.8. This fact further highlights the effectiveness of the conformalisation step is in such a scenario.
Notably, all four models under consideration demonstrate resilience to misspecification. There is no significant difference in performance when the DGP introduced in this section is modelled with incorrect autoregressive terms, such as AR(1) and AR(3).
Real-world models are often imperfectly specified due to the complexity of economic systems and data limitations, especially when predicting aggregate data like GDP. By intentionally testing models that deviate from the true underlying process, we aim to evaluate the sensitivity of quantile forecasts to such misspecifications. Detailed results can be found in Appendix \ref{appendix:A.2}.

\section{Empirical illustration}
\label{sec:realcase}

In this section, we analyse the performance of conformalised and non-conformalised methods when applied to real-world data. The dataset consists of quarterly values of NFCI and GDP growth in the US from 1973 to 2016, downloaded from the supplementary material of \cite{adrian_vulnerable_2019}. We use their framework and findings as a benchmark for assessing our models' performances. \cite{adrian_vulnerable_2019} centres on predicting GDP growth using an autoregressive model that incorporates the NFCI, which captures both financial and macroeconomic conditions. One of their key findings is that a decline in the NFCI (indicating deteriorating economic conditions) not only shifts the mean of the GDP growth distribution to the left, as one might expect, but also increases the leftward skewness of the distribution. This heightened skewness suggests a greater probability of extreme negative GDP growth outcomes. Interestingly, when the NFCI improves, the same degree of rightward skewness is not observed. This asymmetry underscores how downside risks to GDP growth tend to intensify more than upside risks over time.
Scenarios like this, where the behaviour of the target variable distribution, or specific quantiles within it, is of interest, provide an ideal context to assess the effectiveness of conformalisation. In these cases, calibration becomes a critical metric, ensuring that the predicted probabilities across all quantiles accurately reflect the true distribution of future GDP growth.

It is important to note a methodological difference between how \cite{adrian_vulnerable_2019} measured calibration and how it was measured in our simulation study, even though this difference does not lead to any practical discrepancies in the results. In our simulation study, we calculate the unconditional empirical coverage \(\hat{P}(\hat{Q}_i)\), which is simply the proportion of test data that falls below the empirical quantile value \(\hat{Q}_i\) estimated by our model. In contrast, \cite{adrian_vulnerable_2019} employed the Probability Integral Transform (PIT).
Specifically, if \(X\) is a random variable with a continuous cumulative distribution function \(F_X(x)\), then the PIT is defined as:
\[
U = F_X(X)
\]
Here, \(U\) is a new random variable that, under the assumption that \(X\) follows the distribution described by \(F_X\), is uniformly distributed on the interval \([0, 1]\).
When applying the PIT to a set of observations, the transformed values resulting should follow a uniform distribution if the assumed model (represented by \(F_X\)) is correct. 
Another key difference from our approach in our earlier simulation study is that \cite{adrian_vulnerable_2019} retrain their model at each time step, incorporating new information to predict the next observation. 
To ensure the most direct comparison between the performance of our models and those in the original study, we adopt both the PIT approach and the procedure of retraining the model at each time step in this section.

\subsection{Quantile Regression}
We begin by focusing on QR model, which is the same model selected by \cite{adrian_vulnerable_2019} in their study.
In the final step of their estimation process, \citeauthor{adrian_vulnerable_2019} fit a parametric distribution to the estimated quantiles to construct a continuous distribution of future GDP growth. However, because a final smoothing procedure would compromise the properties of CQR methods, we don't implement any smoothing but estimate a much finer grid of quantiles instead. An important consequence is that models which interpolate a continuous distribution tend to be better calibrated at the tails of the distribution compared to models that rely solely on estimating a large number of discrete quantiles without smoothing. This smoothness allows the model to provide more accurate and realistic probability estimates for extreme events, specifically at our two most extreme quantiles: below 0.01 and above 0.99.

\begin{figure}[H]
    \centering
    \subfloat[\label{}]{
        \includegraphics[scale=0.34]{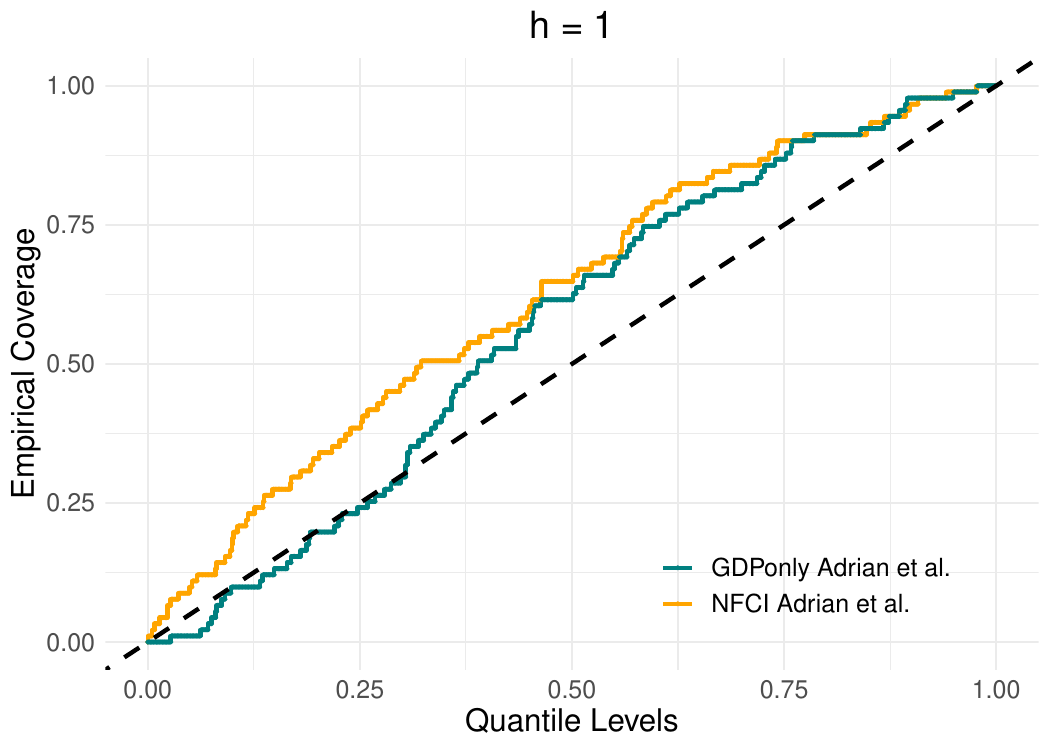}
    }
    \hspace{1em}
    \subfloat[\label{}]{
        \includegraphics[scale=0.34]{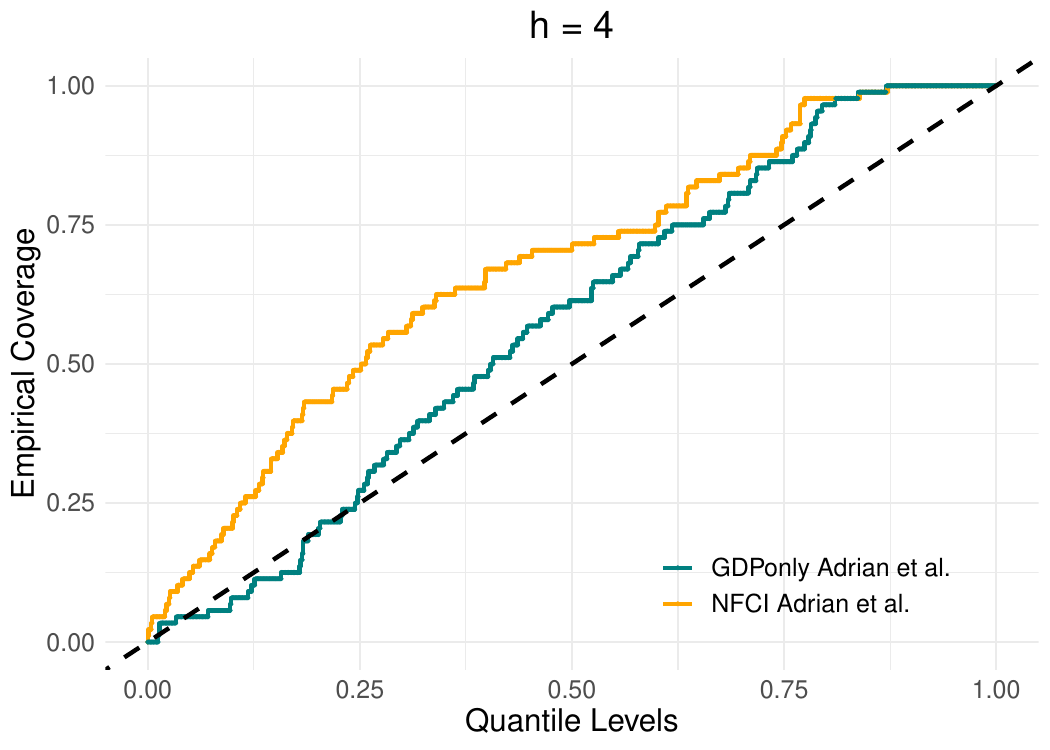}
    }
    \captionsetup{font=footnotesize}
    \caption[]{The calibration curves are from QR models that use either the NFCI combined with the previous GDP growth value as predictors, or only the previous GDP growth value as the sole predictor. The values \(h = 1\) and \(h = 4\) correspond to one-quarter-ahead and one-year-ahead estimates of GDP growth, respectively. 
    
    The same plot can be found in Panel C and Panel D in Figure 11 in \cite{adrian_vulnerable_2019}}.
    \label{fig:figure6}
\end{figure}

In Figure \ref{fig:figure6}, we present the calibration curves obtained by \cite{adrian_vulnerable_2019} The values \(h = 1\) and \(h = 4\) correspond to one-quarter-ahead and one-year-ahead estimates of GDP growth, respectively. These estimates are generated by QR models that use either the NFCI combined with the previous GDP growth value as predictors, or only the previous GDP growth value as the sole predictor. As shown, there is room for improving calibration. Our attention focuses only on the model using NFCI.

\begin{figure}[H]
    \centering
    \subfloat[\label{}]{
        \includegraphics[scale=0.34]{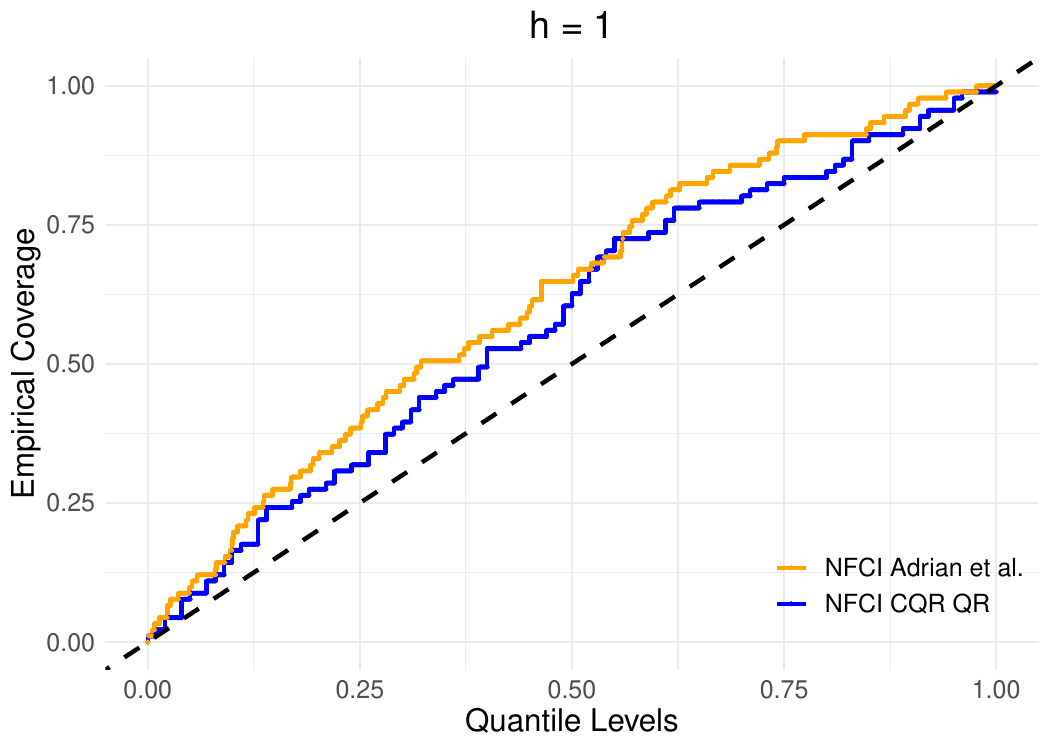}
    }
    \hspace{1em}
    \subfloat[\label{}]{
        \includegraphics[scale=0.34]{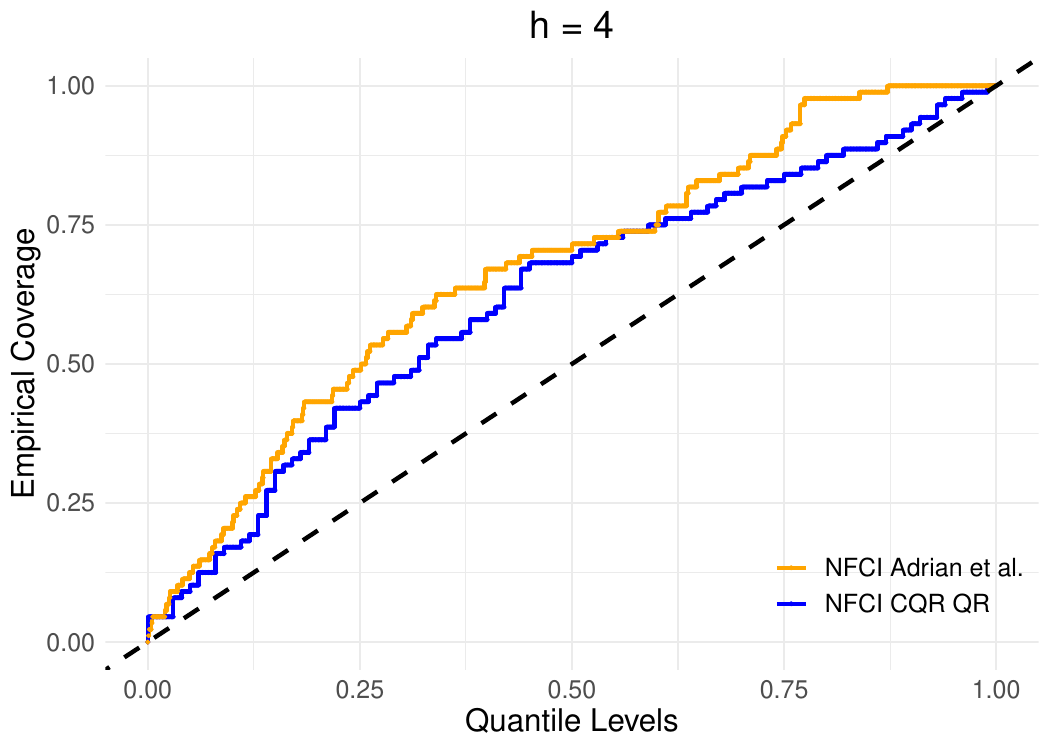}
    }
    \captionsetup{font=footnotesize}
    \caption[]{NFCI \citeauthor{adrian_vulnerable_2019} refers to the original work. NFCI CQR QR is our estimate applying the CQR QR algorithm to the same data. The values \(h = 1\) and \(h = 4\) correspond to one-quarter-ahead and one-year-ahead estimates of GDP growth, respectively.}
    \label{fig:figure7}
\end{figure}

To begin, we immediately apply our CQR algorithm and compare it to the version used by \cite{adrian_vulnerable_2019}
The results show that CQR QR model provides a noticeable, though not particularly significant, improvement in calibration (Figure \ref{fig:figure7}). 

\begin{figure}[H]
    \centering
    \subfloat[\label{}]{
        \includegraphics[scale=0.34]{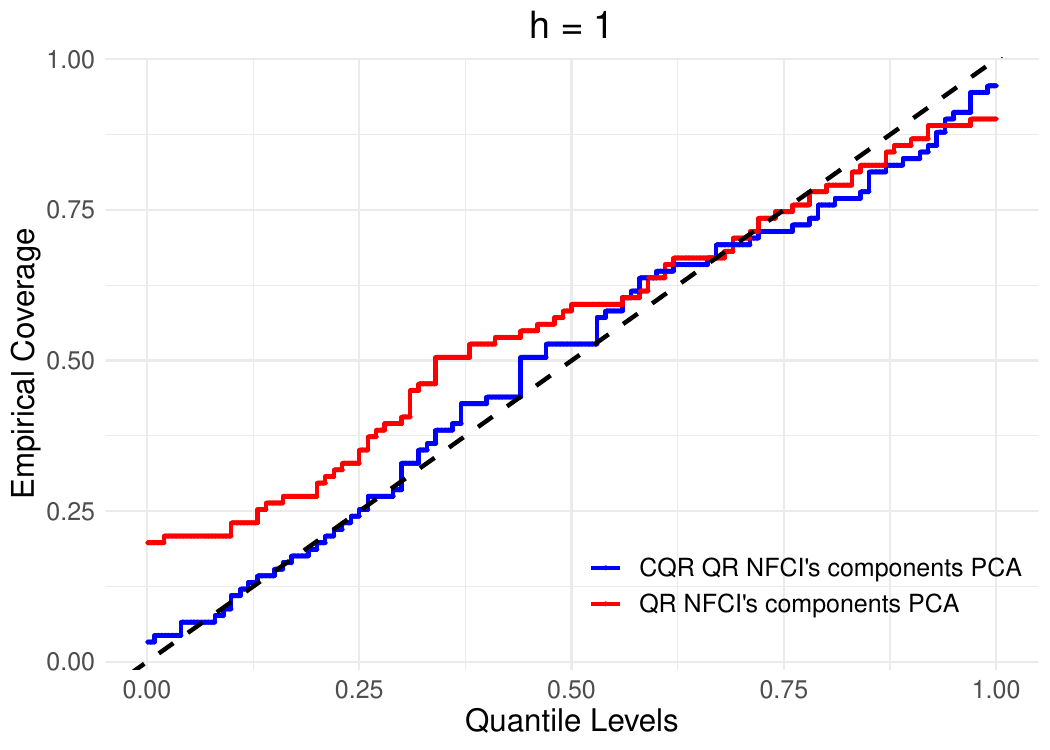}
    }
    \hspace{1em}
    \subfloat[\label{}]{
        \includegraphics[scale=0.34]{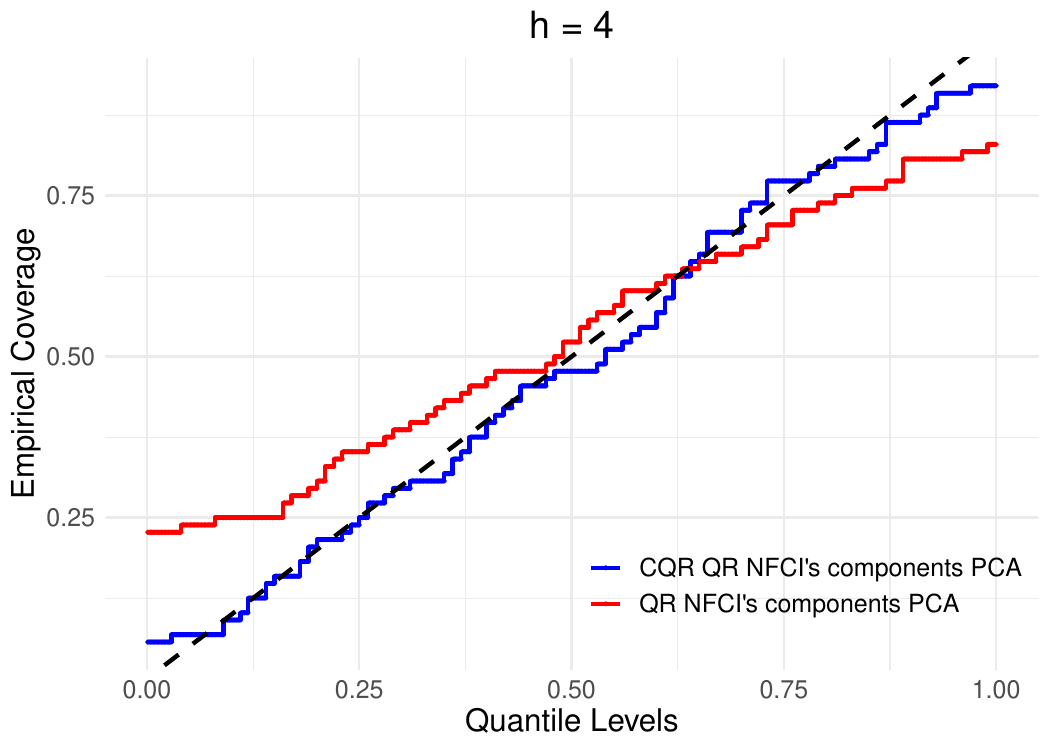}
    }
    \captionsetup{font=footnotesize}
    \caption[]{PCA components of NFCI are used in QR and CQR QR models as predictors for GDP growth.
    
    The values \(h = 1\) and \(h = 4\) correspond to one-quarter-ahead and one-year-ahead estimates of GDP growth, respectively.}
    \label{fig:figure8}
\end{figure}

Simulations suggest that the CQR algorithm performs very well when estimating with a significant number of covariates. The NFCI is an index that tracks financial markets by aggregating around 100 different data series and indexes. Given how effective CQR QR is at handling a large number of covariates, and considering that using individual components can provide more detailed information for predicting future GDP growth than a single aggregated index, we have removed the NFCI from the model and have instead used its individual components.
Since the number of covariates now exceeds the length of the time series, we have performed a Principal Component Analysis (PCA), retaining the components that explain 90\% of the total variance.
The CQR algorithm is now much better calibrated (Figure \ref{fig:figure8}). We also observe that the calibration curves for the QR model tend to flatten at the extreme quantiles, where calibration is poor. A similar issue was observed during the simulation with exogenous variables (Figure \ref{fig:figure3}).

\begin{figure}[H]
    \centering
    \subfloat[\label{}]{
        \includegraphics[scale=0.34]{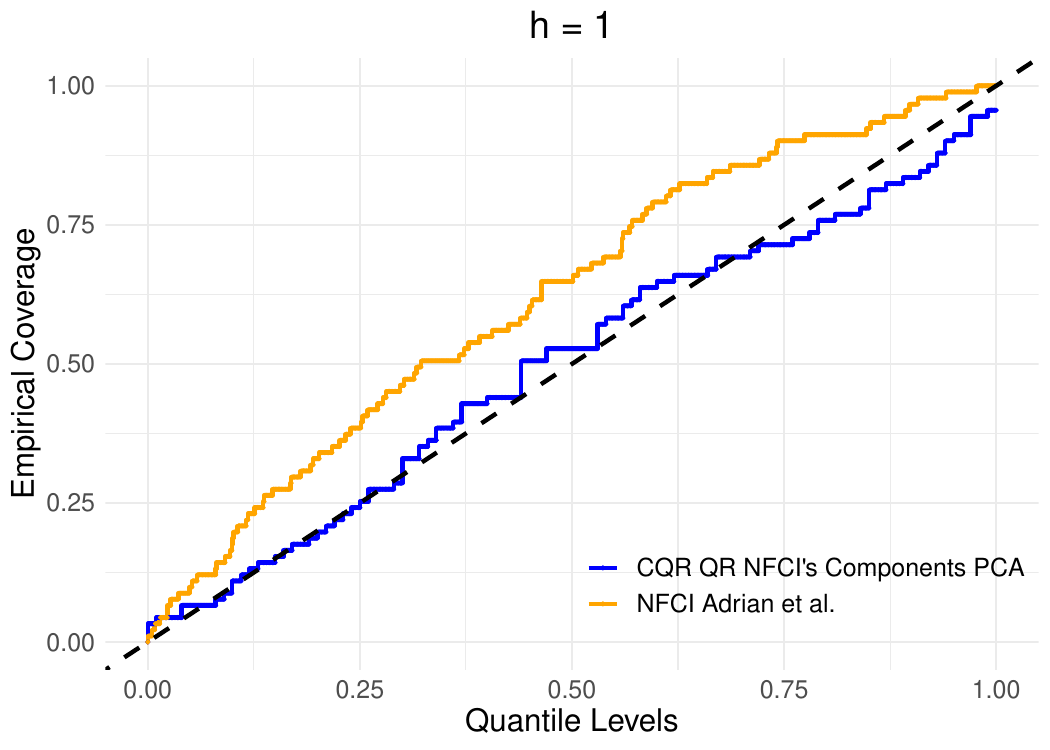}
    }
    \hspace{1em}
    \subfloat[\label{}]{
        \includegraphics[scale=0.34]{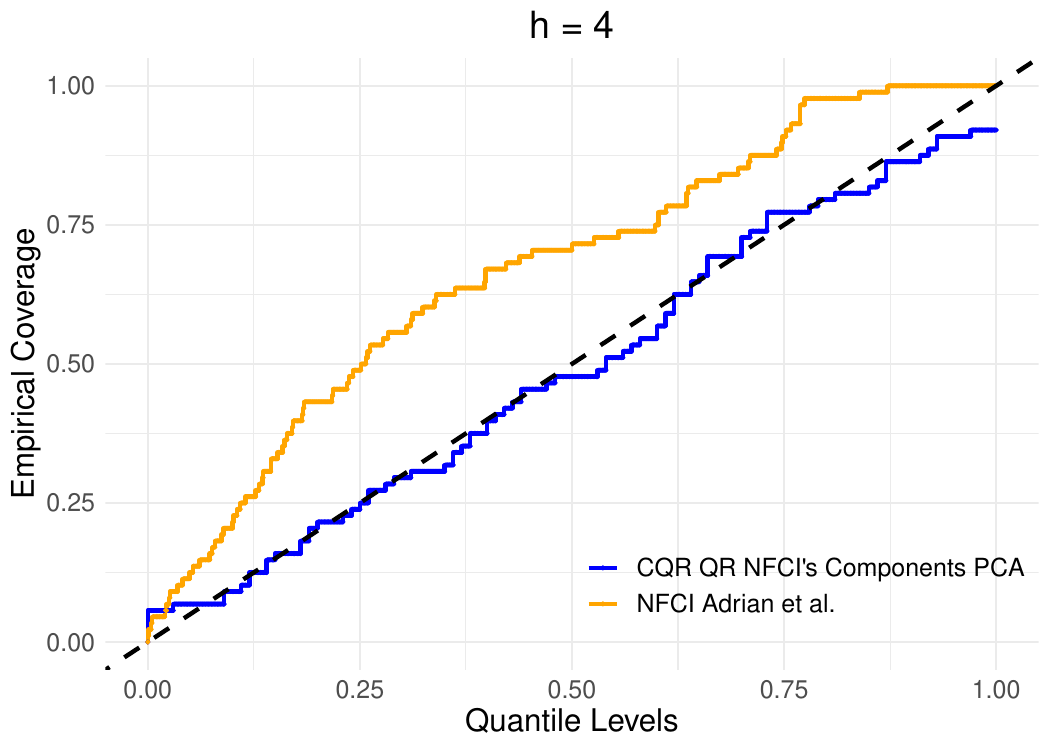}
    }
    \captionsetup{font=footnotesize}
    \caption[]{Comparing curves from Figure \ref{fig:figure6} and \ref{fig:figure8}}.
    \label{fig:figure9}
\end{figure}

Lastly, in Figure \ref{fig:figure9} we compare the calibration of the original model by \cite{adrian_vulnerable_2019} and that of the CQR QR model with the components of NFCI.

By also examining the MAE results in Table \ref{table:table5}, we can draw several conclusions. Using the NFCI components instead of the aggregated index enables the QR and CQR QR models to better capture the distribution of future GDP growth. However, although the QR model reduces its MAE when using NFCI components instead of NFCI itself, its inability to correctly calibrate the extreme quantiles highlights the importance of the calibration step.

\begin{table}[H]
    \caption*{\textbf{}}
    \centering 
    \begin{tabular}{|p{18em} c c|}
    \hline
    \rowcolor{lavender}
    \textbf{MAE} & \textbf{QR} &  \textbf{CQR QR} \\
    \hline \hline
    \textbf{NFCI h = 1} & 0.125  & 0.075 \\
    \textbf{NFCI h = 4} & 0.177  & 0.115 \\
    \textbf{PCA on NFCI components h = 1} & 0.075 & 0.028  \\
    \textbf{PCA on NFCI components h = 4} & 0.079  & 0.023 \\
    \hline
    \end{tabular}
    \\[10pt]
    \captionsetup{font=footnotesize}
    \caption{MAE (\ref{eq:MAE}) of all the models and time steps \(h\). NFCI QR is the original work of \cite{adrian_vulnerable_2019}}
    \label{table:table5}
\end{table}

\subsection{Quantile Random Forest}

We now switch our focus to QRF and CQR QRF. Since the original work by \cite{adrian_vulnerable_2019} does not include QRF, we constructed both QRF and CQR QRF models to predict future GDP growth using the NFCI and past growth values (Figure \ref{fig:figure9}). Our findings indicate that the conformalisation procedure significantly improves calibration, particularly for \(h = 1\).

\begin{figure}[H]
    \centering
    \subfloat[\label{}]{
        \includegraphics[scale=0.34]{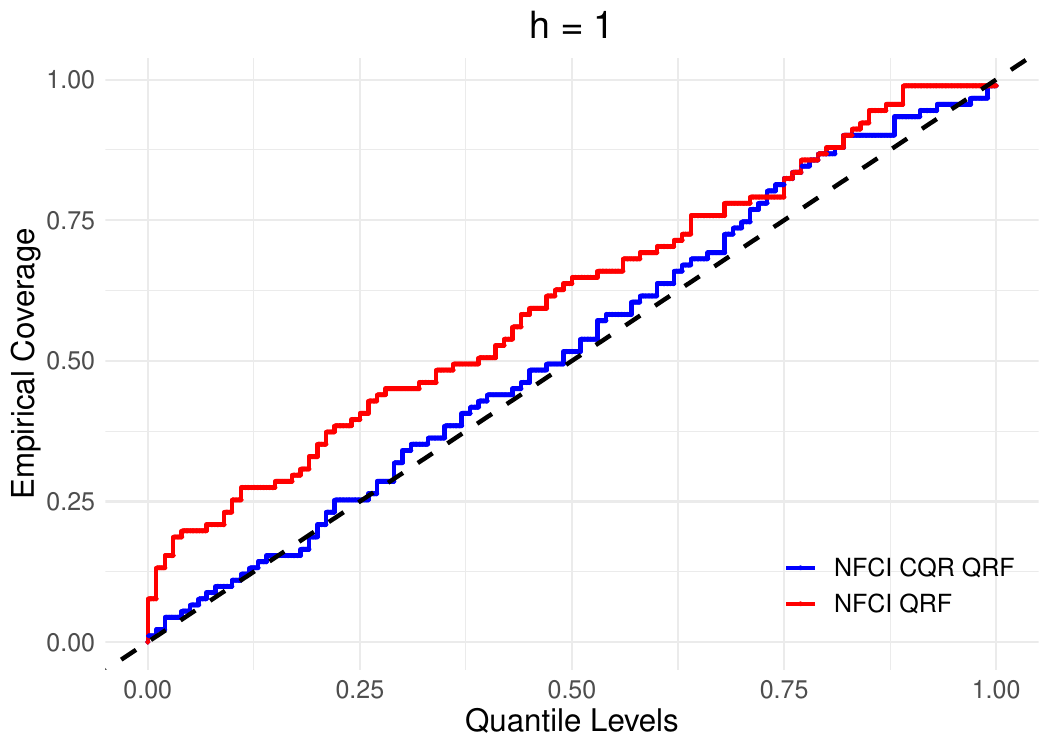}
    }
    \hspace{1em}
    \subfloat[\label{}]{
        \includegraphics[scale=0.34]{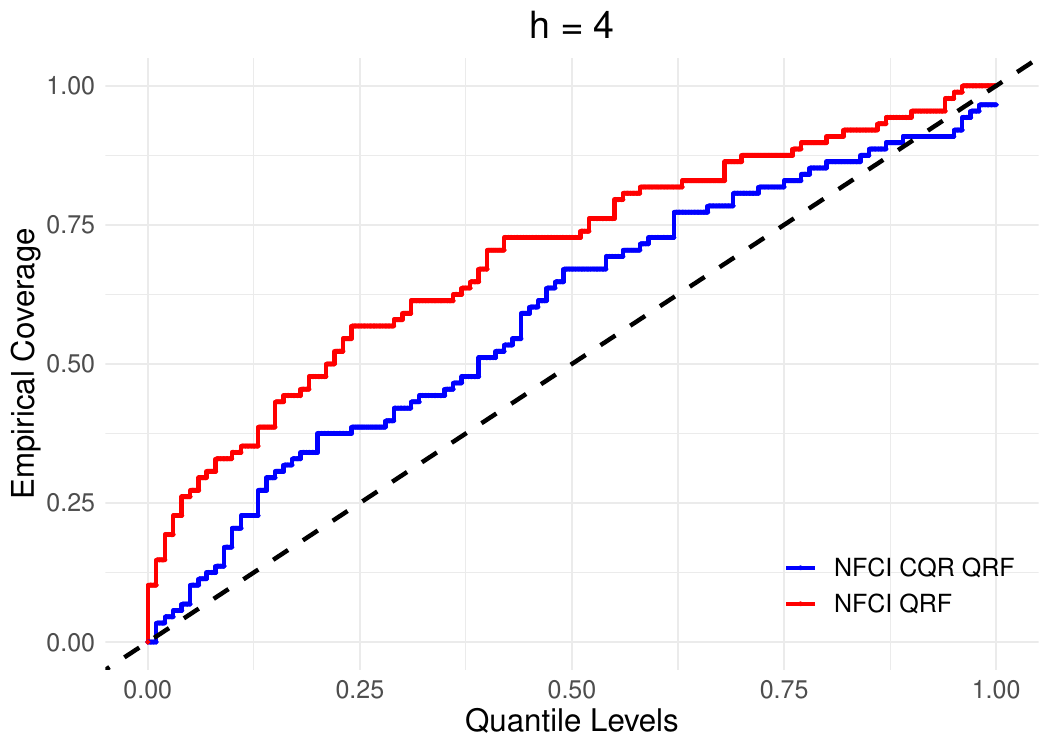}
    }
    \captionsetup{font=footnotesize}
    \caption[]{QRF and CQR QRF calibration curves using NFCI combined with the previous GDP growth value as predictors. The values \(h = 1\) and \(h = 4\) correspond to one-quarter-ahead and one-year-ahead estimates of GDP growth, respectively.}.
    \label{fig:figure10}
\end{figure}

\begin{figure}[H]
    \centering
    \subfloat[\label{}]{
        \includegraphics[scale=0.34]{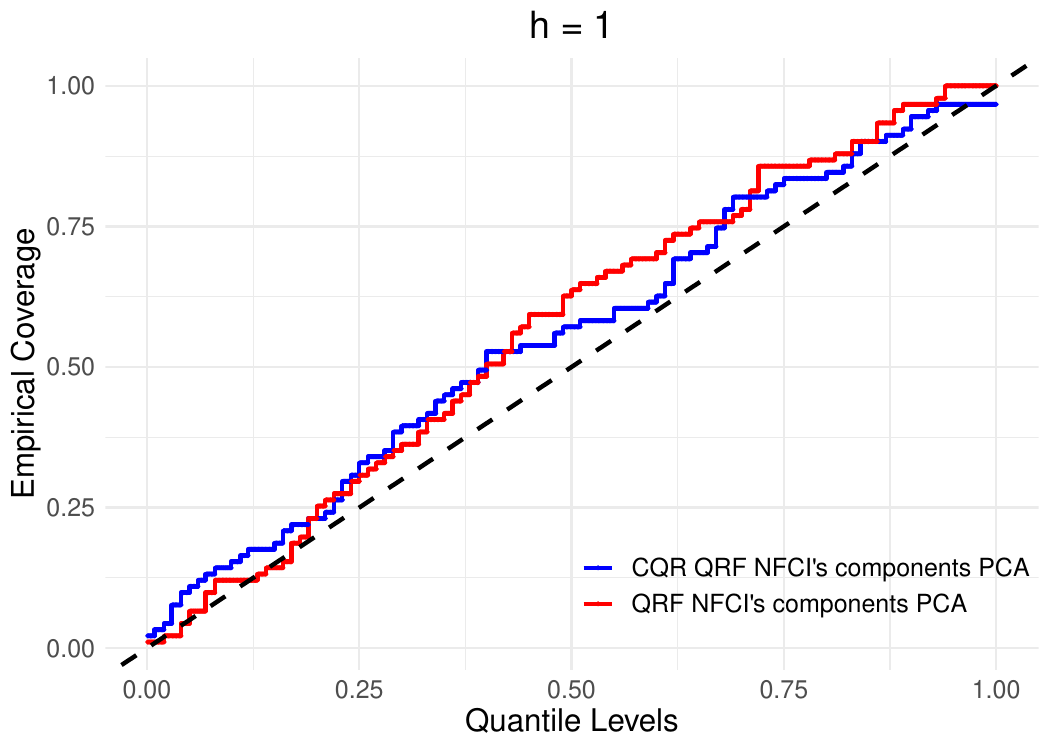}
    }
    \hspace{1em}
    \subfloat[\label{}]{
        \includegraphics[scale=0.34]{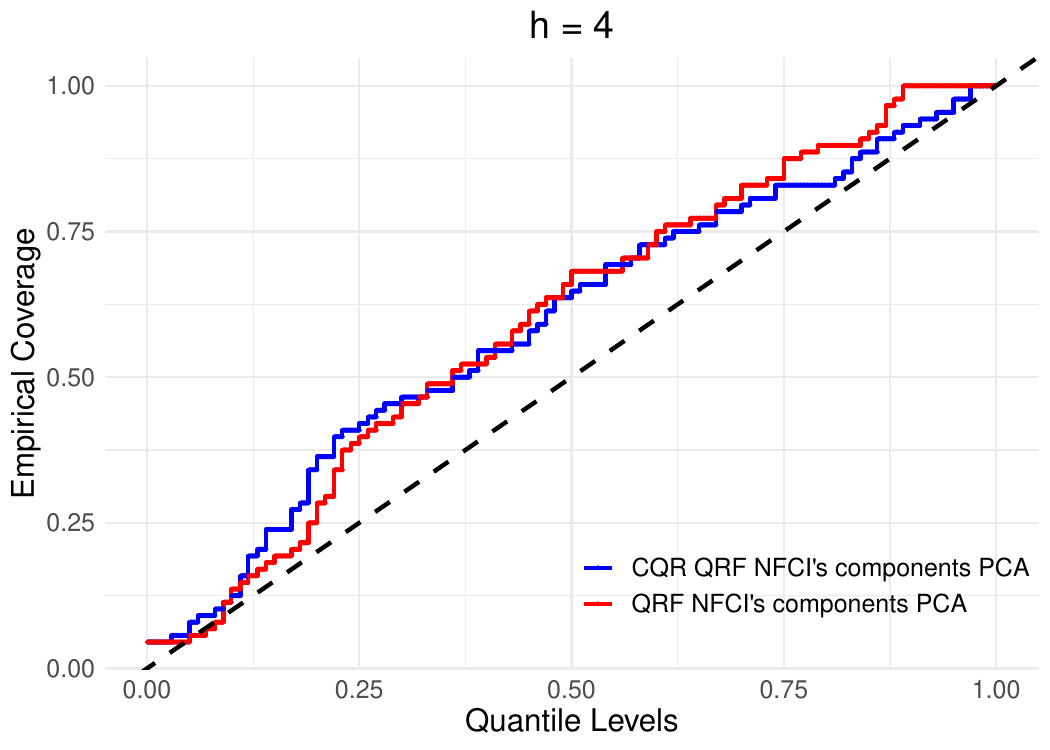}
    }
    \captionsetup{font=footnotesize}
    \caption[]{The NFCI is broken down into its components, a PCA is performed, retaining the principal components capturing 90\% of total variance. Then they are used in QRF and CQR QRF models as predictors for GDP growth.
    
    The values \(h = 1\) and \(h = 4\) correspond to one-quarter-ahead and one-year-ahead estimates of GDP growth, respectively.}.
    \label{fig:figure11}
\end{figure}

We proceed as detailed in the previous subsection, removing NFCI and using its components. Based on our simulation study, we anticipated the calibration of CQR QRF would not be significantly better than that of QRF when disaggregating NFCI into its components, as QRF alone is already effective at handling a significant number of covariates (Figure \ref{fig:figure5}).
The results of QRF and CQR QRF with NFCI divided into its components are presented in Figure \ref{fig:figure11}. CQR only slightly improves the performances of the standard method, and the two models are largely equivalent. Furthermore, the benefits of using NFCI's components are less evident this time; for example, at \(h=1\), using the aggregated NFCI itself appears to be preferable.

\begin{table}[H]
    \caption*{\textbf{}}
    \centering 
    \begin{tabular}{|p{18em} c c|}
    \hline
    \rowcolor{lavender}
    \textbf{MAE} & \textbf{QRF} &  \textbf{CQR QRF} \\
    \hline \hline
    \textbf{NFCI h = 1} & 0.104  & 0.017 \\
    \textbf{NFCI h = 4} & 0.192  & 0.091 \\
    \textbf{PCA on NFCI components h = 1} & 0.064 & 0.051 \\
    \textbf{PCA on NFCI components h = 4} & 0.094  & 0.086 \\
    \hline
    \end{tabular}
    \\[10pt]
    \captionsetup{font=footnotesize}
    \caption{MAE (\ref{eq:MAE}) of all the models and time steps \(h\).}
    \label{table:table6}
\end{table}

\section{Conclusion}

In this paper, we explore the application of an adaptation of Conformalised Quantile Regression in the context of economic forecasting, particularly for computing GaR. The primary goal of the work was to assess the empirical performance of conformalised versus non-conformalised quantile prediction models and to determine whether conformalisation enhances the reliability and robustness of quantile estimates.

Our results from extensive simulation studies and the empirical illustration provide several key insights. Firstly, we find that  QR and QRF models offer complementary strengths in different scenarios. When one model performs poorly, the other often compensates by providing better results. For example, QR performs better than QRF in scenarios with data drawn from distributions without finite moments, such as the Cauchy distribution. In contrast, QRF shows superior performance in high-dimensional environments. This complementarity suggests that the combined use of both methods, possibly through model averaging or ensemble techniques, could provide a more robust approach to quantile estimation across diverse data contexts, even without conformalisation.

The simulation study demonstrated that CQR methods offer substantial, consistent improvements in calibration compared to standard QR and QRF models. Even in contexts where QR performs well, its conformalised counterpart consistently matched or exceeded its calibration performance. The only exception is the nearly unit root time series, where all models—conformalised or not—struggled to provide reliable estimates. In such cases, the inherent persistence and (near-)random-walk behaviour of the series leads to poor model performance. This highlights a limitation of the discussed quantile estimation techniques in handling strongly persistent processes.

In the empirical illustration, focused on predicting US GDP growth using financial and macroeconomic data, CQR methods demonstrate their value by improving accuracy and calibration.
Our analysis also highlights that when calibration is poor, the primary issues often arise at extreme quantile levels, which are of the greatest interest, e.g. for GaR analysis. Even models that exhibit a relatively low MAE can still be problematic if they fail to accurately estimate these extreme quantiles. Since these quantiles are crucial for assessing the risk of adverse economic outcomes, poor calibration in these areas undermines the reliability of the model's predictions and limits its utility for GaR analysis and hence policymaking. However, CQR methods demonstrate a clear advantage in this regard by consistently providing better calibration at extreme quantiles.

Our results demonstrate that both conformalised and non-conformalised models are remarkably robust to misspecifications in the data generating process, such as incorrect autoregressive terms or variations in error distributions. This resilience is particularly relevant in real-world applications where the true underlying structure of the data is often unknown, reinforcing the practical utility of these methods in economic forecasting and risk management. More research is needed to fully understand the extent of this robustness across various settings and to explore potential losses in model performance under different forms of misspecification.

A limitation of our findings arises from the assumption of exchangeability, which is not only crucial for general CP methods but is also central to the coverage guarantee (\ref{eq:main_eq}) of our specific approach. 
In the simulated scenarios, CQR methods exhibit calibration properties that closely align with theoretical coverage guarantees. This alignment is particularly important because it directly supports the ability to control the probability of extreme adverse outcomes in future GDP growth. By ensuring that the probability of exceeding a specified threshold is kept within a desired risk limit, our approach enables economic policymakers, such as central banks, to set and manage risk levels effectively. 
However, in the empirical illustration, there is no evidence that the coverage property holds, as indicated by calibration curves that frequently lie above the diagonal line, reflecting a tendency to overestimate coverage rather than underestimate it. The time dependence inherent in the GDP growth rate series significantly violates the exchangeability assumption, compromising the coverage guarantee. Despite this, the performance of CQR methods remained strong, often exceeding that of the standard methods.
    Addressing the challenges posed by non-exchangeable data or time series in CP is an area already explored in the literature, as in \cite{barber_conformal_2023}.

Overall, this paper contributes to the literature by providing the first application of CP methods to quantile estimation, offering a novel perspective on how these methods can be adapted and utilized for economic and financial forecasting. Our findings underscore the potential of CQR in enhancing the robustness of risk assessments.
Future research could explore methods to overcome the limitations posed by the exchangeability assumption in time-series data, as well as adapt other CP techniques for quantile estimation. Additionally, the proposed method could be applied to other economic and financial contexts, such as Value-at-Risk estimation, and in any setting where accurate quantile estimation is crucial.

%\section*{Acknowledgments}
%I would like to express my sincere gratitude to my advisor, Prof. Simone Vantini, and my co-advisors, Matteo Fontana and Luca Neri, for their invaluable guidance and support throughout this research. I am also deeply thankful to my family for their unwavering encouragement, patience, and love, which have been a constant source of strength and inspiration during this journey.
\vspace{17em}

\bibliography{references_matteo, references_luca}

\appendix
\section{Appendix A}
\subsection{Calibration curves for the AR(2) Exogenous model}
\label{appendix:A.1}

\begin{figure}[H]
    \centering
    \subfloat[\label{}]{
        \includegraphics[width=0.29\textwidth]{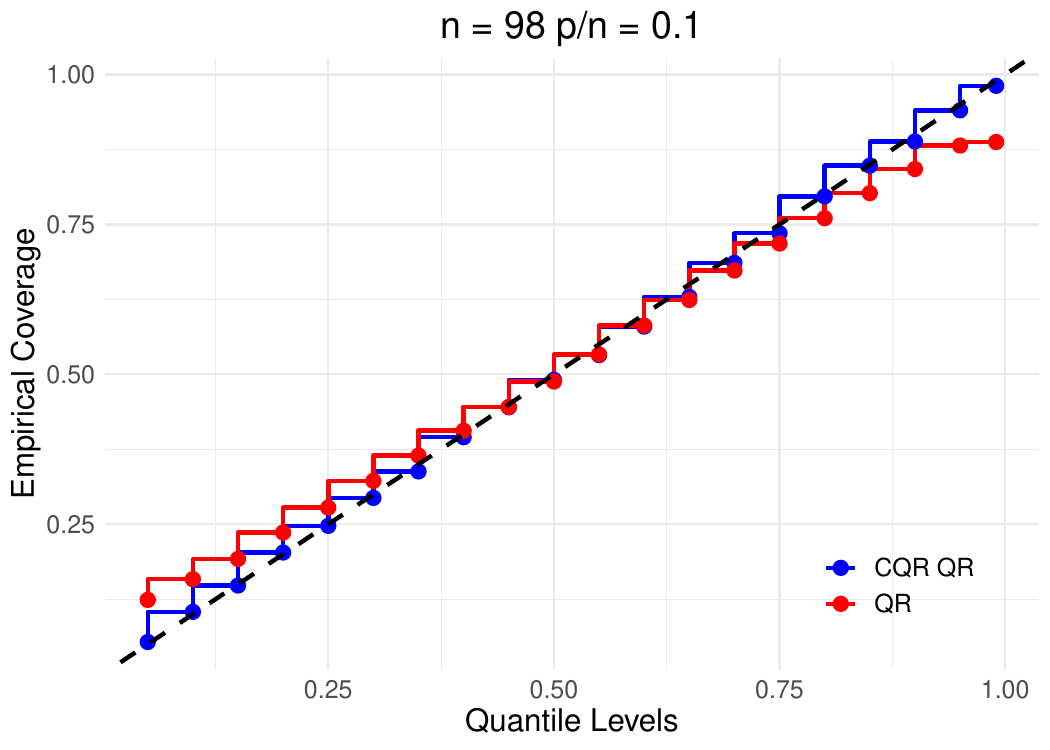}
    }
    \hspace{1em}
    \subfloat[\label{}]{
        \includegraphics[width=0.29\textwidth]{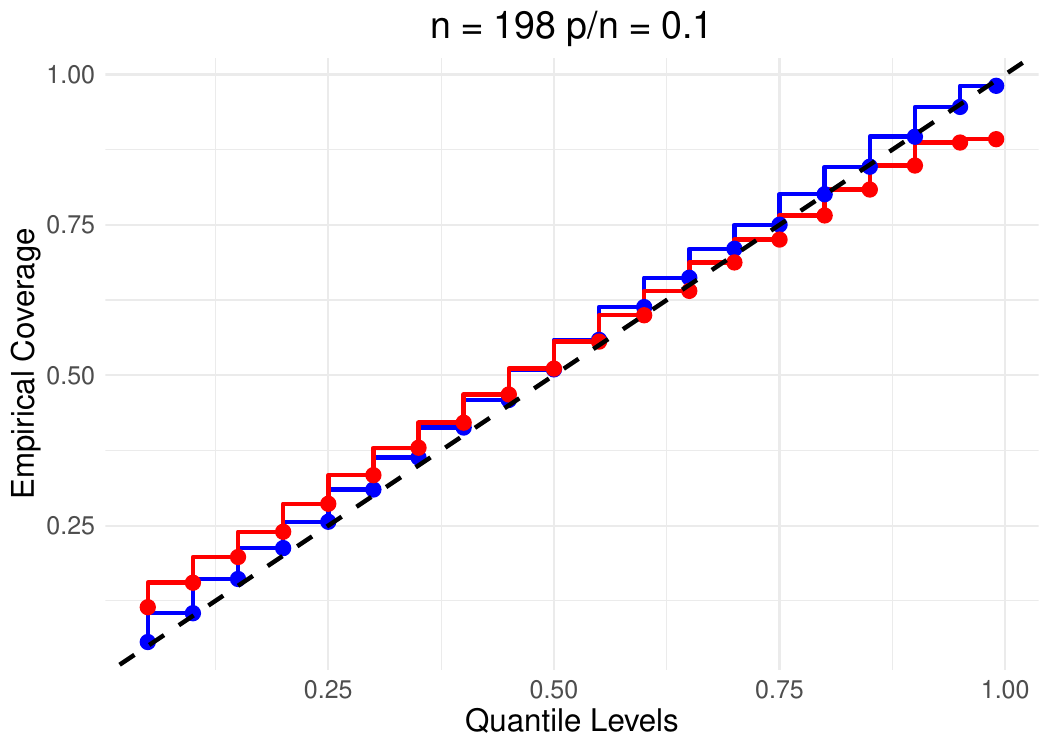}
    }
    \hspace{1em}
    \subfloat[\label{}]{
        \includegraphics[width=0.29\textwidth]{QR_Exogeneus_n998_0.1.pdf}
    }\\
    \vspace{1em}
    \subfloat[\label{}]{
        \includegraphics[width=0.29\textwidth]{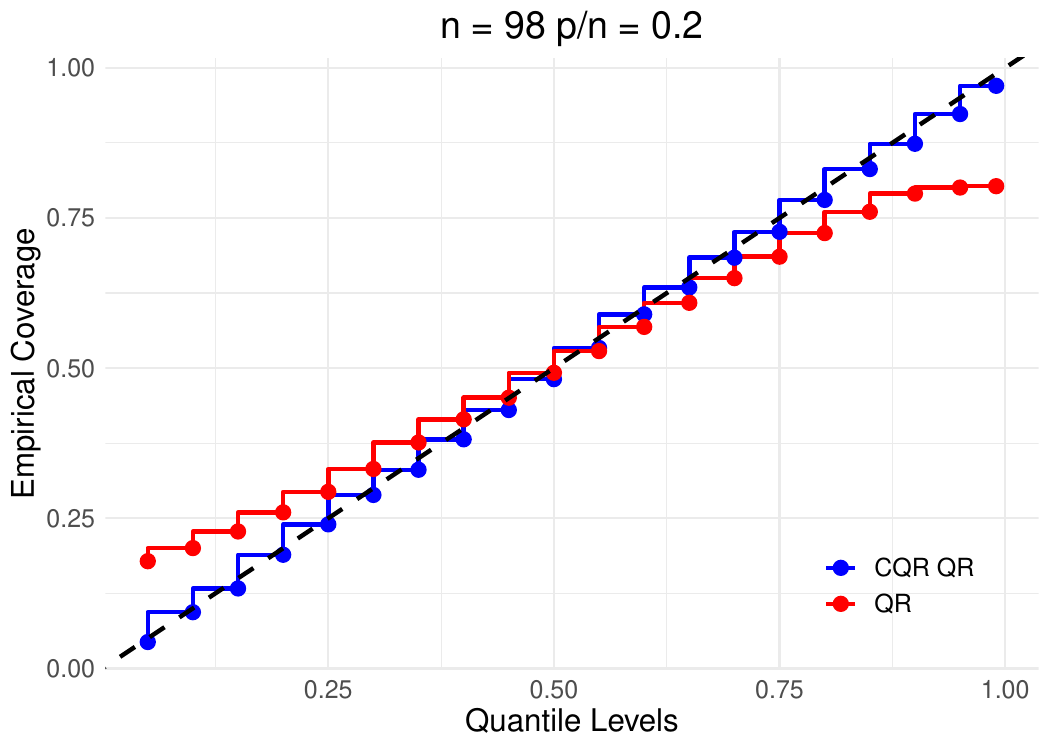}
    }
    \hspace{1em}
    \subfloat[\label{}]{
        \includegraphics[width=0.29\textwidth]{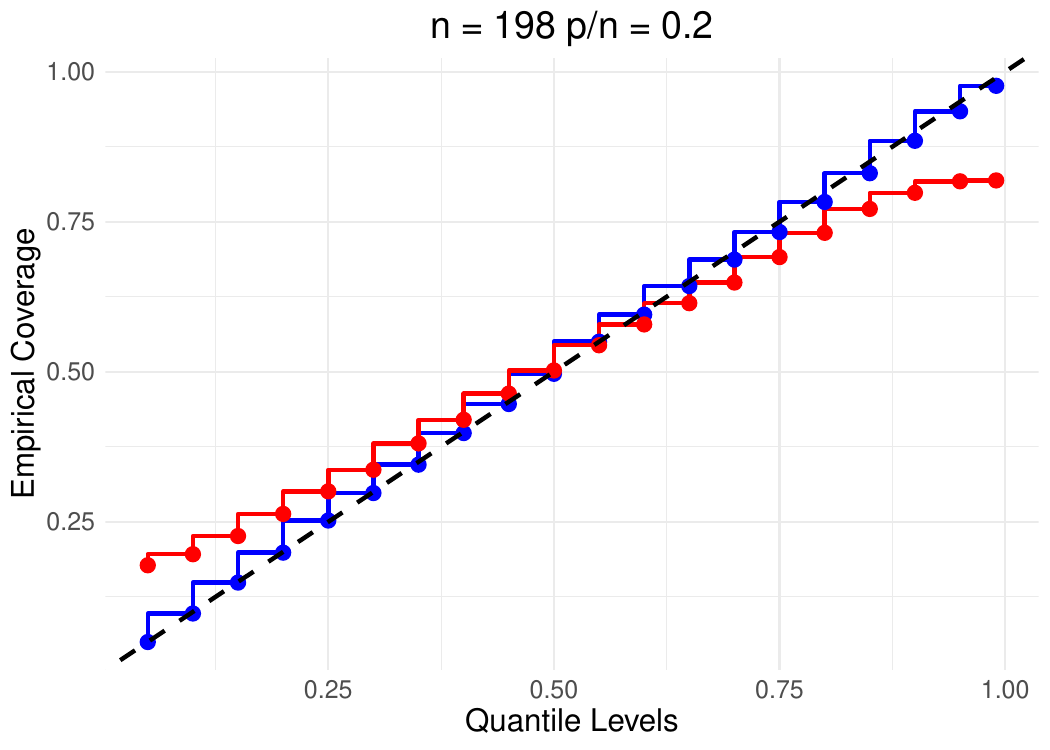}
    }
    \hspace{1em}
    \subfloat[\label{}]{
        \includegraphics[width=0.29\textwidth]{QR_Exogeneus_n998_0.2.pdf}
    }\\
    \vspace{1em}
    \subfloat[\label{}]{
        \includegraphics[width=0.29\textwidth]{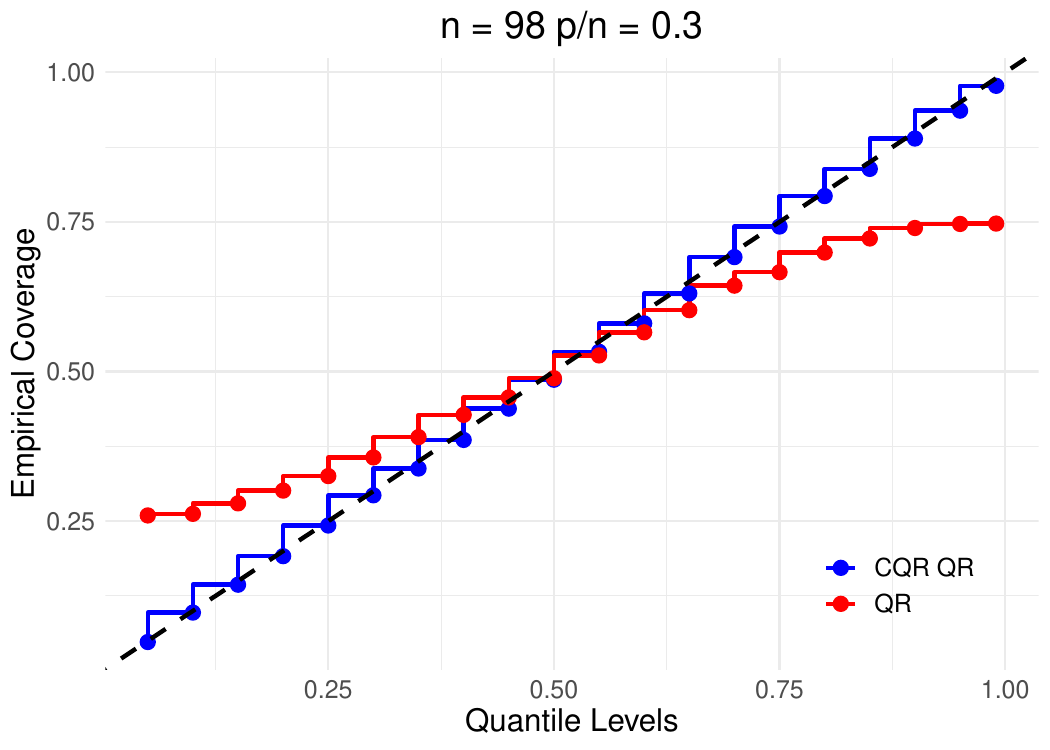}
    }
    \hspace{1em}
    \subfloat[\label{}]{
        \includegraphics[width=0.29\textwidth]{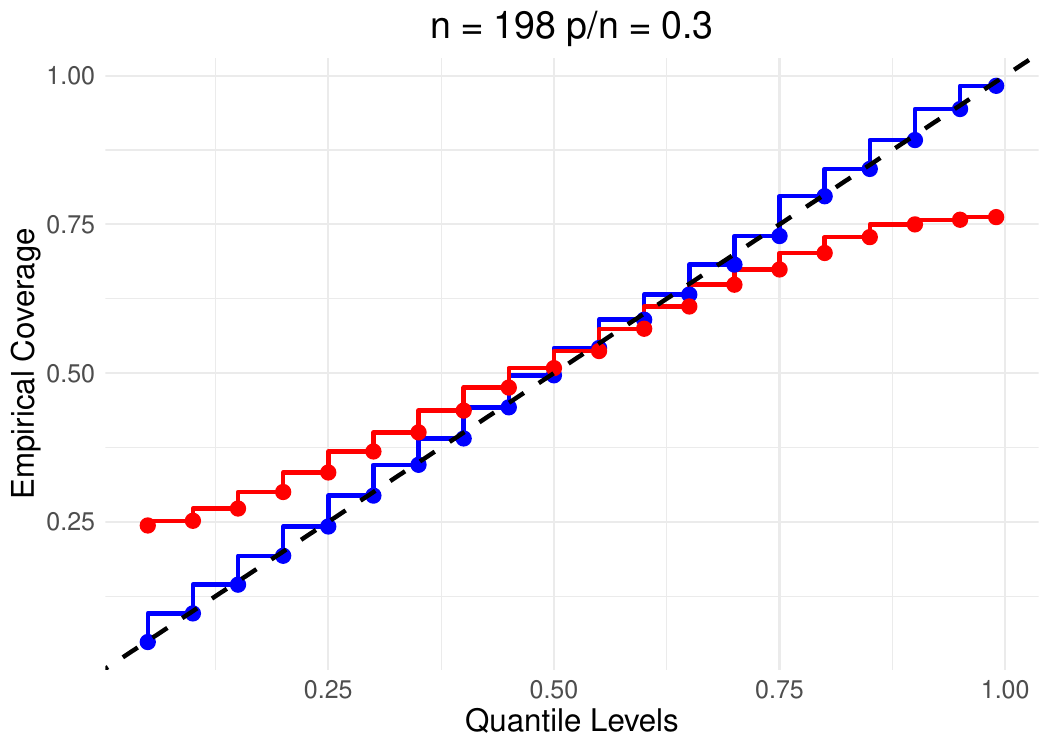}
    }
    \hspace{1em}
    \subfloat[\label{}]{
        \includegraphics[width=0.29\textwidth]{QR_Exogeneus_n198_0.3.pdf}
    }\\
    \vspace{1em}
    \subfloat[\label{}]{
        \includegraphics[width=0.29\textwidth]{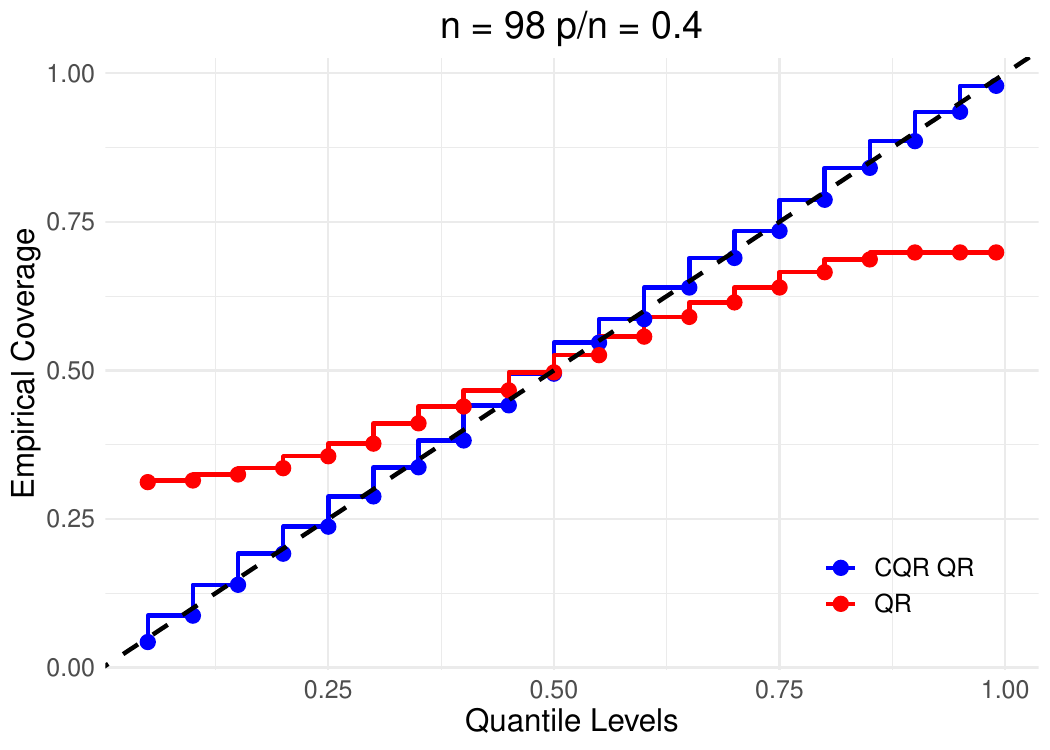}
    }
    \hspace{1em}
    \subfloat[\label{}]{
        \includegraphics[width=0.29\textwidth]{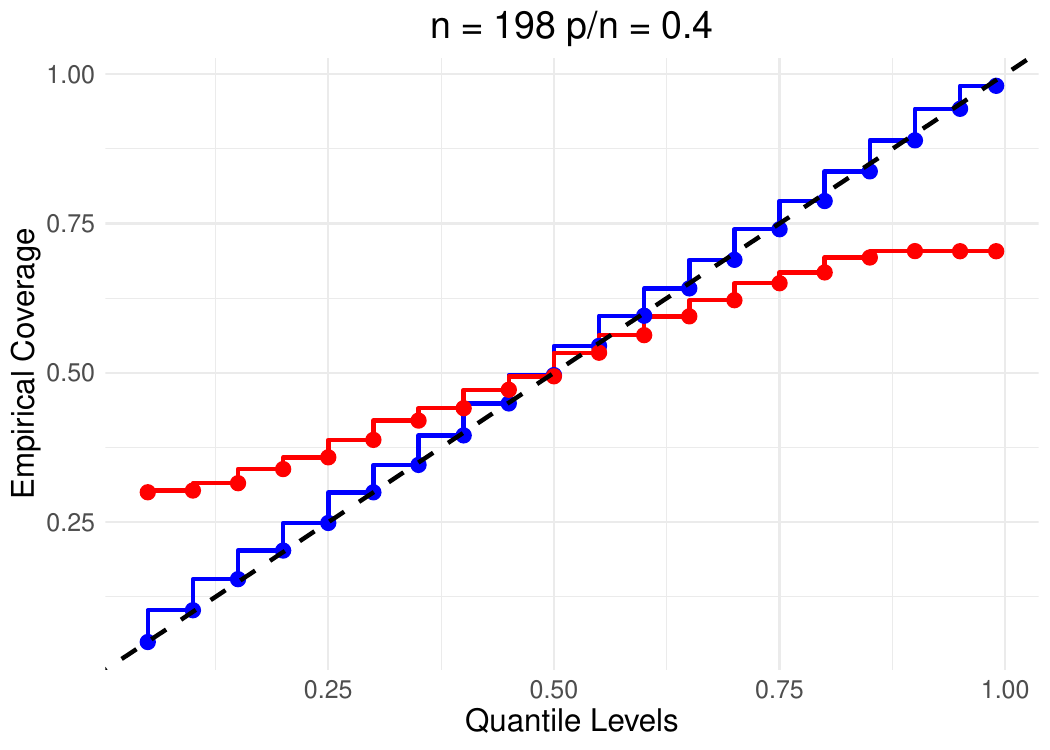}
    }
    \hspace{1em}
    \subfloat[\label{}]{
        \includegraphics[width=0.29\textwidth]{QR_Exogeneus_n998_0.4_.pdf}
    }
    \captionsetup{font=footnotesize}
    \caption[]{Calibration curves for QR and CQR QR for the 3 values of \(n\) and \(p/n\) = 0.1 (first row), 0.2 (second row), 0.3 (third row), and 0.4 (fourth row).}
    \label{fig:figure12}
\end{figure}

\begin{figure}[H]
    \centering
    \subfloat[\label{}]{
        \includegraphics[width=0.29\textwidth]{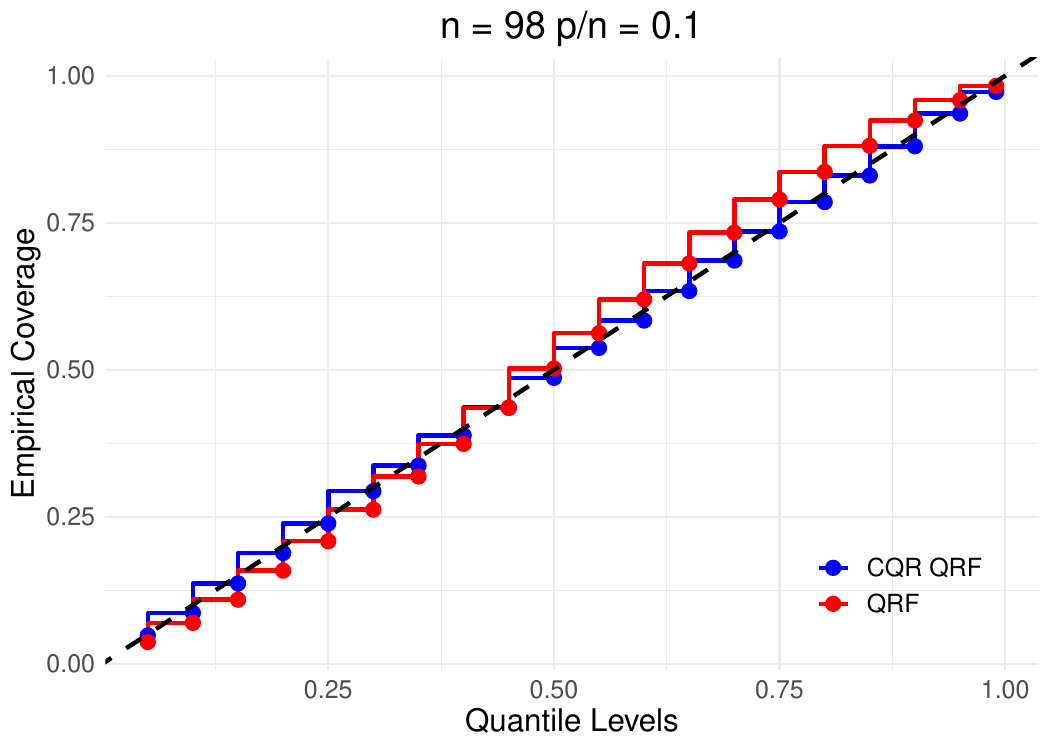}
    }
    \hspace{1em}
    \subfloat[\label{}]{
        \includegraphics[width=0.29\textwidth]{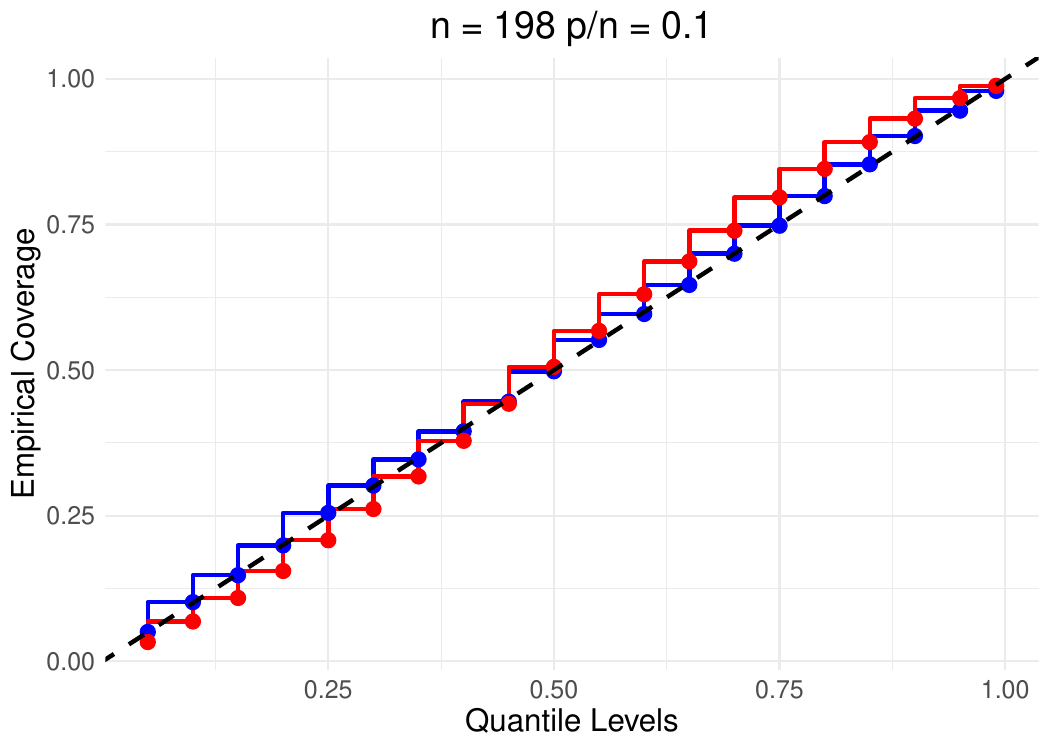}
    }
    \hspace{1em}
    \subfloat[\label{}]{
        \includegraphics[width=0.29\textwidth]{QRF_Exogeneus_n998_0.1.pdf}
    }\\
    \vspace{1em}
    \subfloat[\label{}]{
        \includegraphics[width=0.29\textwidth]{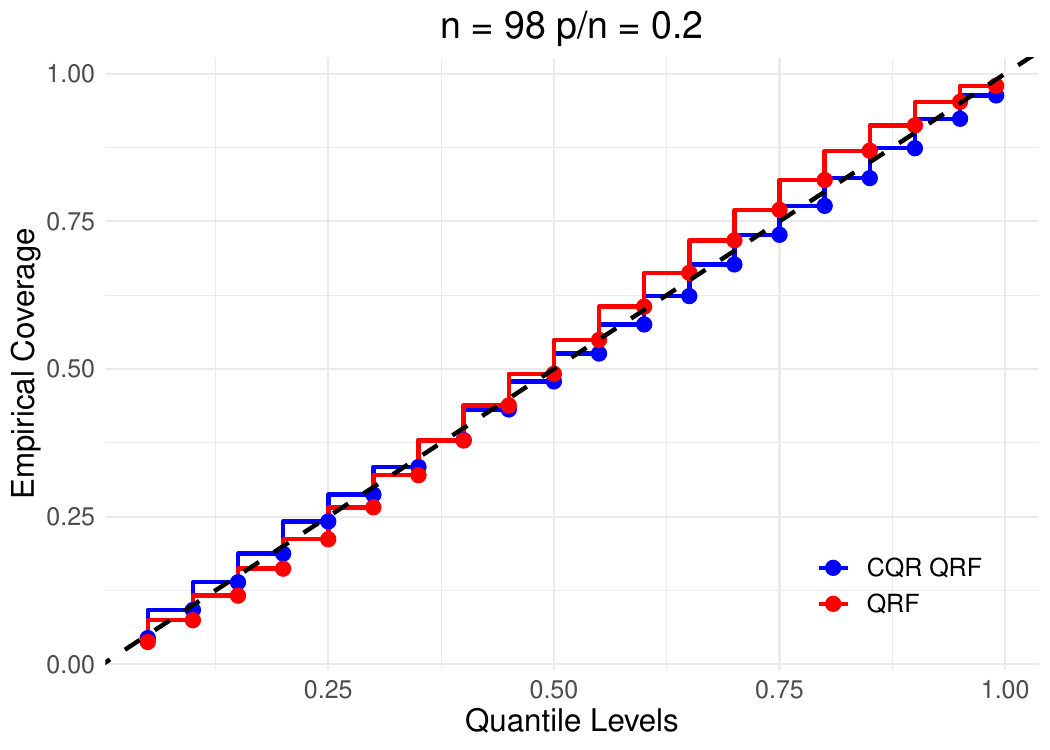}
    }
    \hspace{1em}
    \subfloat[\label{}]{
        \includegraphics[width=0.29\textwidth]{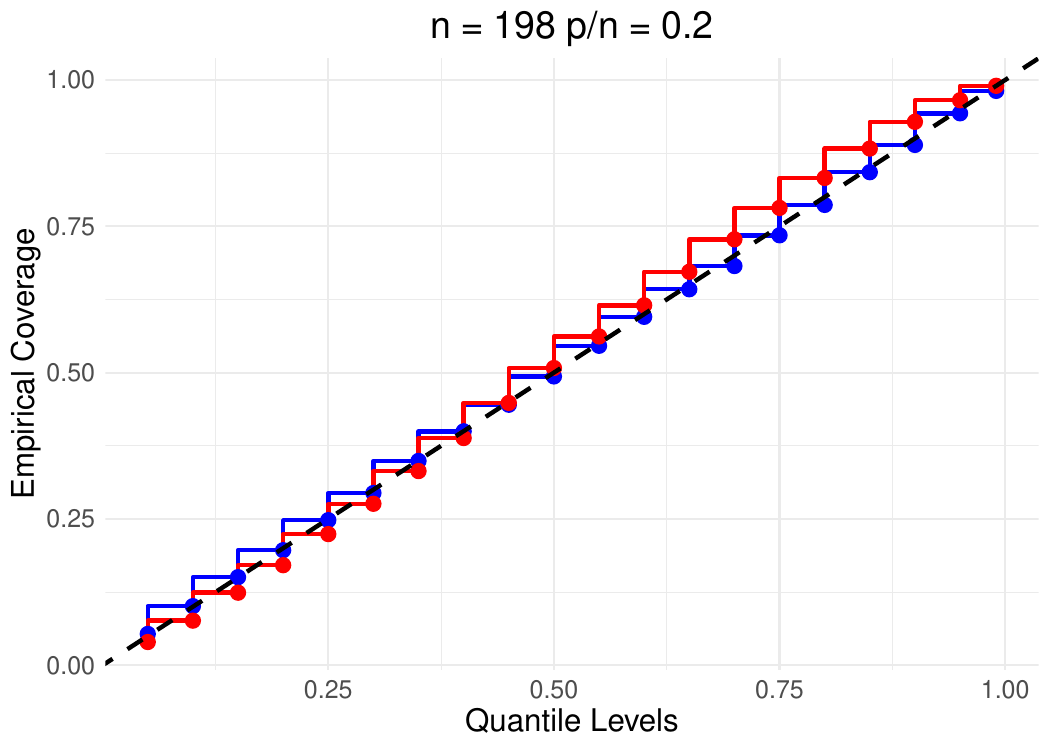}
    }
    \hspace{1em}
    \subfloat[\label{}]{
        \includegraphics[width=0.29\textwidth]{QRF_Exogeneus_n998_0.2.pdf}
    }\\
    \vspace{1em}
    \subfloat[\label{}]{
        \includegraphics[width=0.29\textwidth]{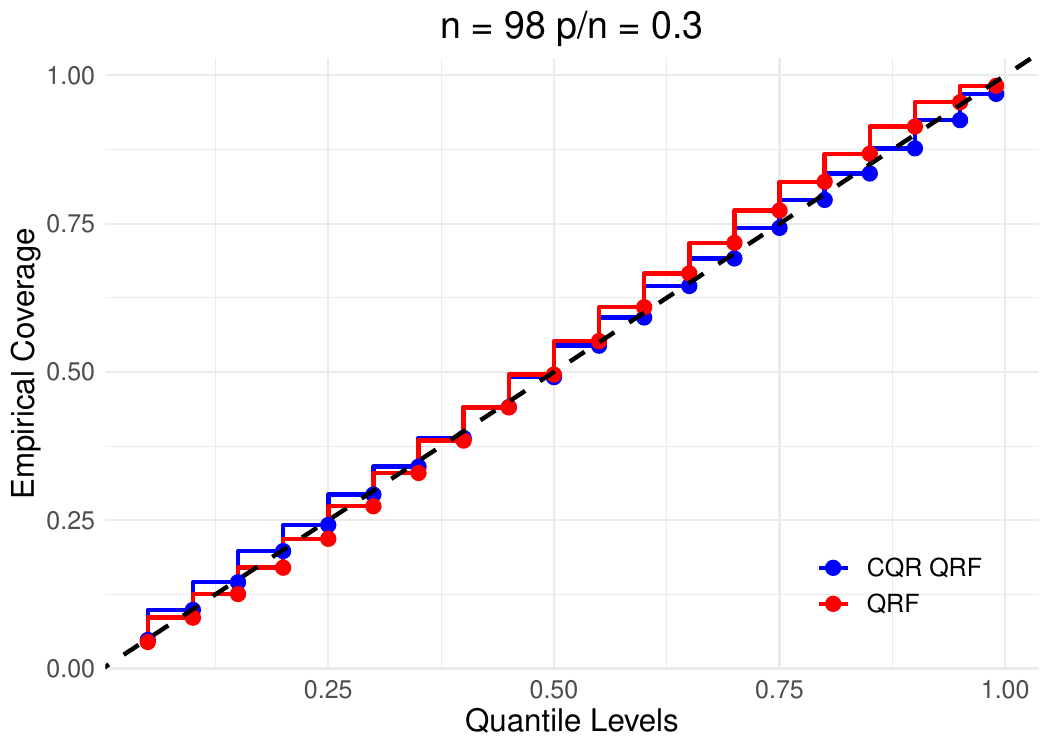}
    }
    \hspace{1em}
    \subfloat[\label{}]{
        \includegraphics[width=0.29\textwidth]{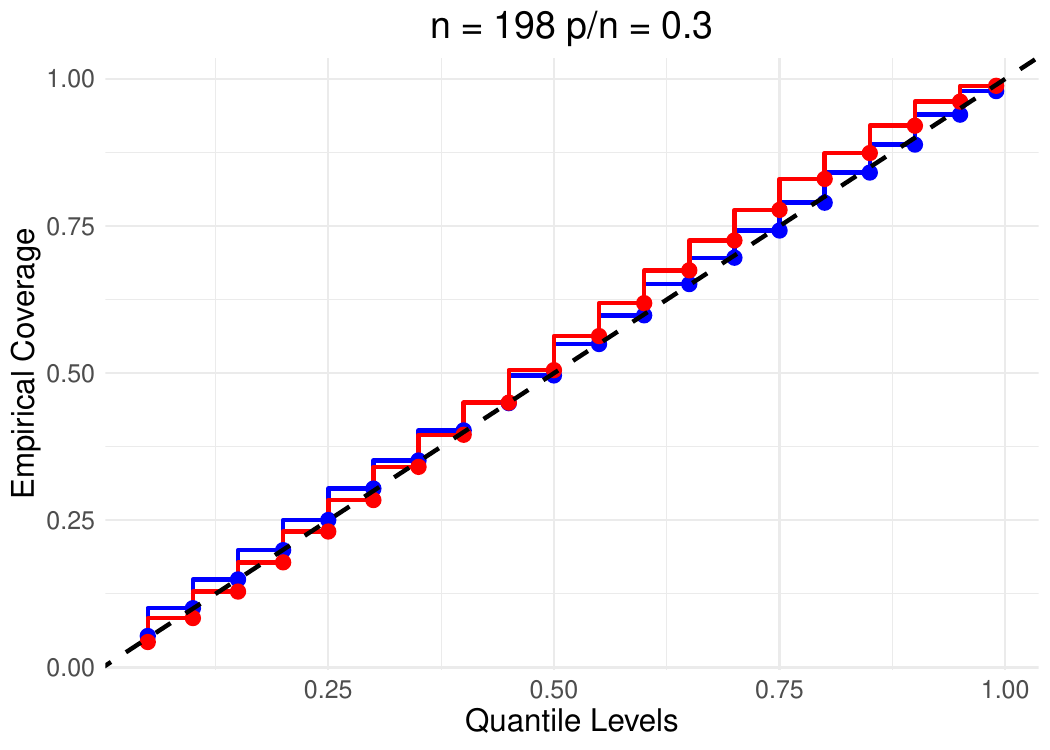}
    }
    \hspace{1em}
    \subfloat[\label{}]{
        \includegraphics[width=0.29\textwidth]{QRF_Exogeneus_n198_0.3.pdf}
    }\\
    \vspace{1em}
    \subfloat[\label{}]{
        \includegraphics[width=0.29\textwidth]{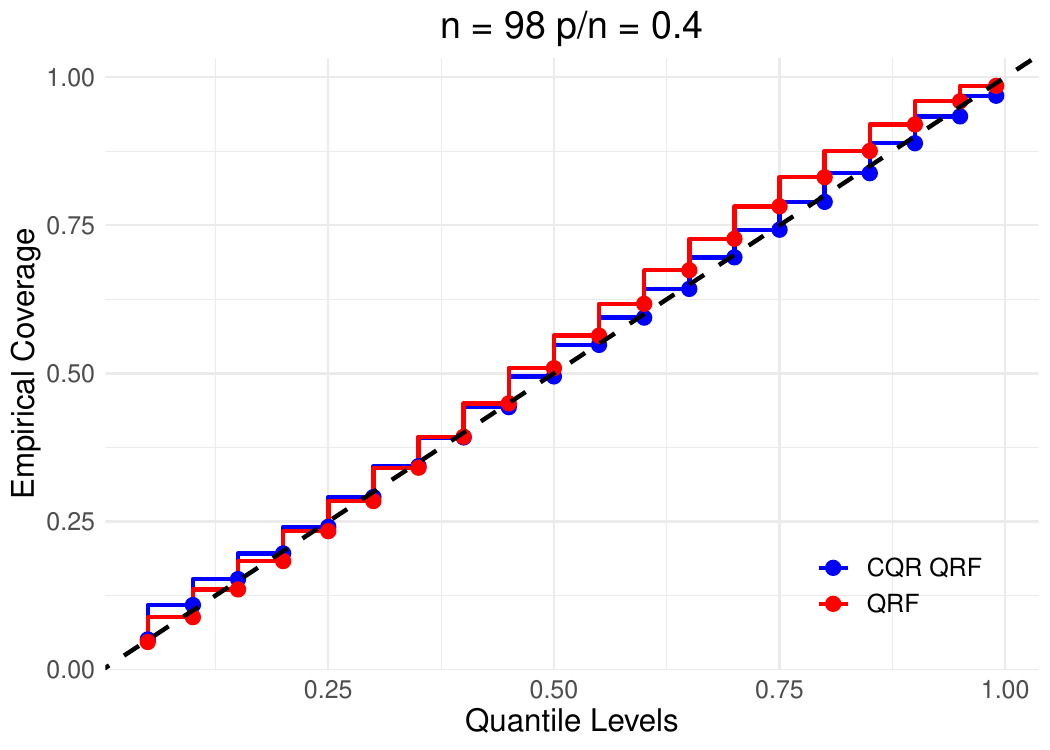}
    }
    \hspace{1em}
    \subfloat[\label{}]{
        \includegraphics[width=0.29\textwidth]{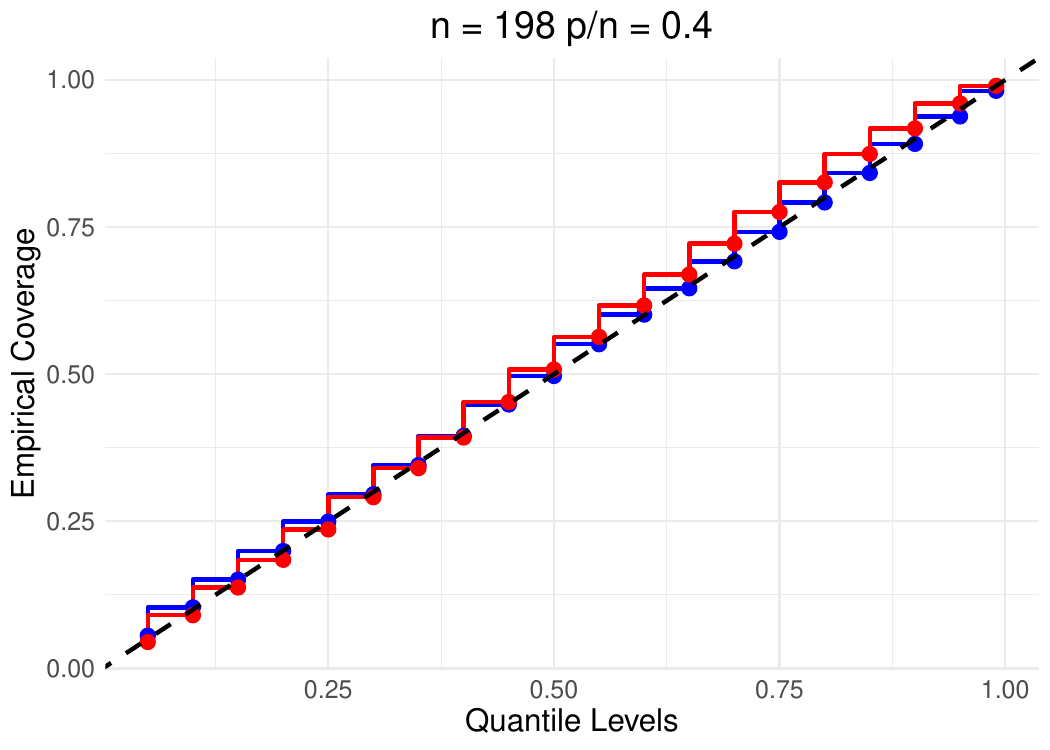}
    }
    \hspace{1em}
    \subfloat[\label{}]{
        \includegraphics[width=0.29\textwidth]{QRF_Exogeneus_n998_0.4.pdf}
    }
    \captionsetup{font=footnotesize}
    \caption[]{Calibration curves for QR and CQR QR for the 3 values of \(n\) and \(p/n\) = 0.1 (first row), 0.2 (second row), 0.3 (third row), and 0.4 (fourth row).}
    \label{fig:figure12}
\end{figure}

\subsection{Resilience to Misspecification for the AR(2) Exogenous model}
\label{appendix:A.2}

We assess the predictive accuracy and robustness of each model by fitting it to data using three different autoregressive specifications: AR(1), AR(2), and AR(3), while keeping the exogenous component correctly specified. By doing so, we intentionally introduce misspecification to determine how well each model can adapt to errors in the assumed data structure and maintain reliable quantile estimates. The AR(1) and AR(3) models represent misspecified alternatives that either simplify or overfit the true AR(2) process, while the correctly specified AR(2) model serves as a benchmark.

In Table \ref{table:table7}, all four models exhibit similar performance across the different autoregressive specifications. This consistency indicates that the models' accuracy and calibration are relatively unaffected by misspecification of the autoregressive component.

\begin{table}
    \caption*{}
    \centering 
    \begin{tabular}{|p{10em} c c c|}
    \hline
    \rowcolor{lavender}
      \textbf{MAE} & \textbf{AR1} & \textbf{AR2} & \textbf{AR3}\\
    \hline \hline
    \textbf{QR} & 0.074  & 0.077   & 0.078  \\
    \textbf{QRF} & 0.018  & 0.018 & 0.018  \\
    \textbf{CQR QR} & 0.006 & 0.008 & 0.007   \\
    \textbf{CQR QRF} & 0.007  & 0.007 & 0.007  \\

    \hline
    \end{tabular}
    \\[10pt]
    \captionsetup{font=footnotesize}
    \caption{MAE (\ref{eq:MAE}) for the 4 models for AR(1), AR(2) and AR(3), averaged on the 4 values of \(p/n\) and on the 3 values of \(n\).}
    \label{table:table7}
\end{table}

Given the resilience of AR(1) and AR(3) models, we further test by modifying the DGP error using \(\epsilon_t \sim \mathcal{N}(0, 1)\) instead of \(\epsilon_t \sim t_2\). A possibility is that a fat-tailed error masks the effect of the autoregressive component, making its impact on the target variable ignorable.
Even with this additional modification, the models' performances are indistinguishable. (Table \ref{table:table8})

\begin{table}[H]
    \caption*{}
    \centering 
    \begin{tabular}{|p{10em} c c c|}
    \hline
    \rowcolor{lavender}
      \textbf{MAE} & \textbf{AR1} & \textbf{AR2} & \textbf{AR3}  \\
    \hline \hline
    \textbf{QR} & 0.075   & 0.077   & 0.079  \\
    \textbf{QRF} & 0.019   & 0.019   & 0.019  \\
    \textbf{CQR QR} & 0.007 & 0.008 & 0.008   \\
    \textbf{CQR QRF} & 0.007 & 0.007 & 0.007  \\

    \hline
    \end{tabular}
    \\[10pt]
    \captionsetup{font=footnotesize}
    \caption{MAE (\ref{eq:MAE}) of the new DGP with normal error, for the 4 models for AR(1), AR(2) and AR(3), averaged on the 4 values of \(p/n\) and on the 3 values of \(n\).}
    \label{table:table8}
\end{table}

\section{Appendix B}
\label{appendix:B}
\subsection{AR(1) with nearly unit root coefficients}
\label{subsec:unitroot}The following DGP is implemented:
\[
Y_t = \phi_1  Y_{t-1} + \epsilon_t
\]
where
\[
 \phi_1 \in \left\{0.95, 1, 1.05\right\}, \quad \epsilon_t \sim \mathcal{N}(0,1).
\]
Financial and economic time series frequently exhibit behaviours close to a unit root ($\phi_1 \approx 1$) \cite{stock_testing_1988}. These characteristics pose significant challenges for traditional forecasting models. In this section, we explore the potential benefits of CQR in addressing these challenges.
Through 3 values of $\phi_1$, we examine different scenarios:
\begin{itemize}
\item$\phi_1 < 1$ (Stationary): Shocks have temporary effects; the series reverts to a mean over time.

\item$\phi_1 = 1$ (Unit Root): Shocks have permanent effects; the series is non-stationary and follows a random walk.

\item$\phi_1 > 1$ (Explosive): Shocks amplify over time; the series exhibits exponential growth or decay.
\end{itemize}

\begin{figure}[H]
    \centering
    \subfloat[\label{}]{        \includegraphics[width=0.29\textwidth]{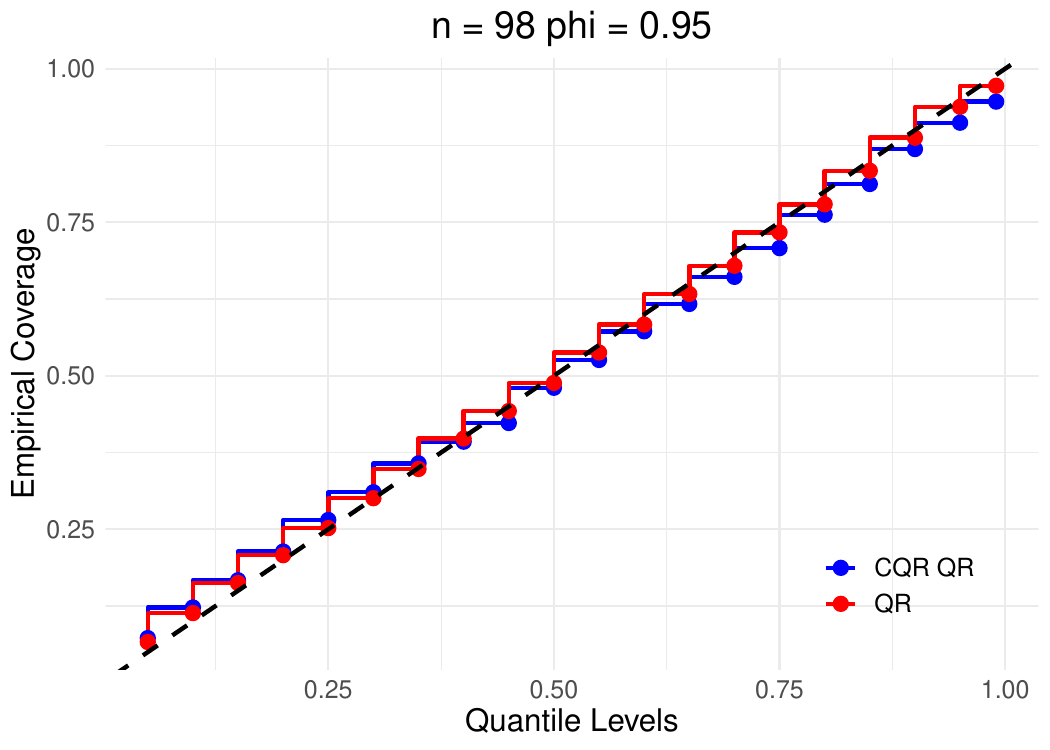}
    }
    \hspace{1em}
    \subfloat[\label{}]{        \includegraphics[width=0.29\textwidth]{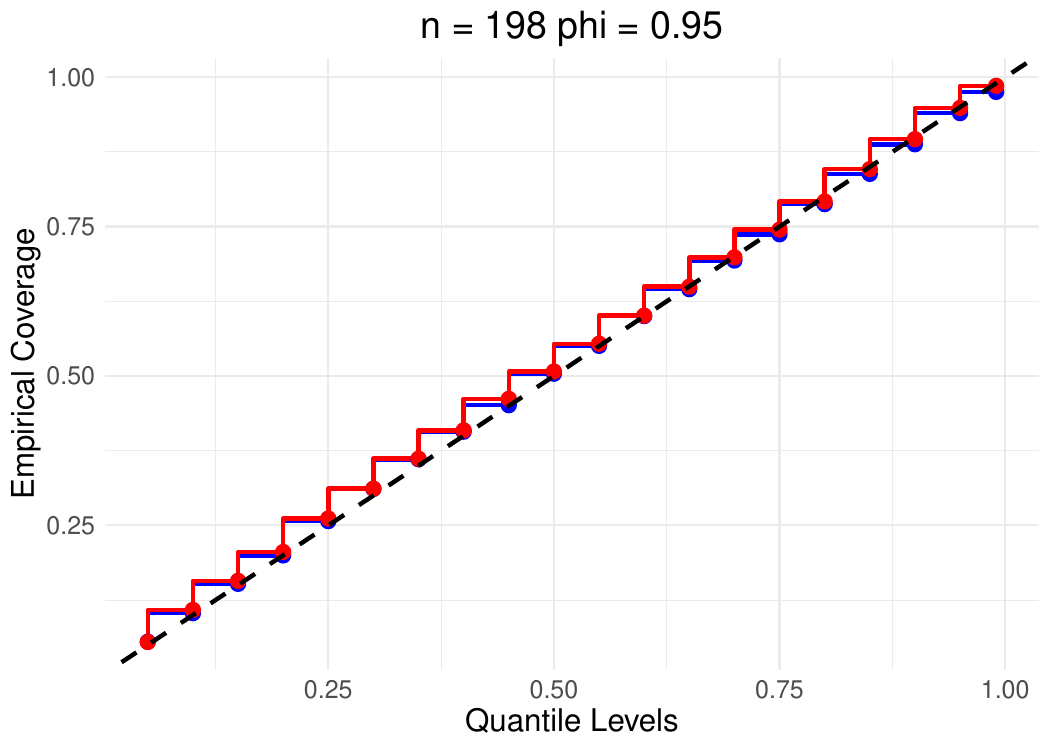}
    }
    \hspace{1em}
    \subfloat[\label{}]{        \includegraphics[width=0.29\textwidth]{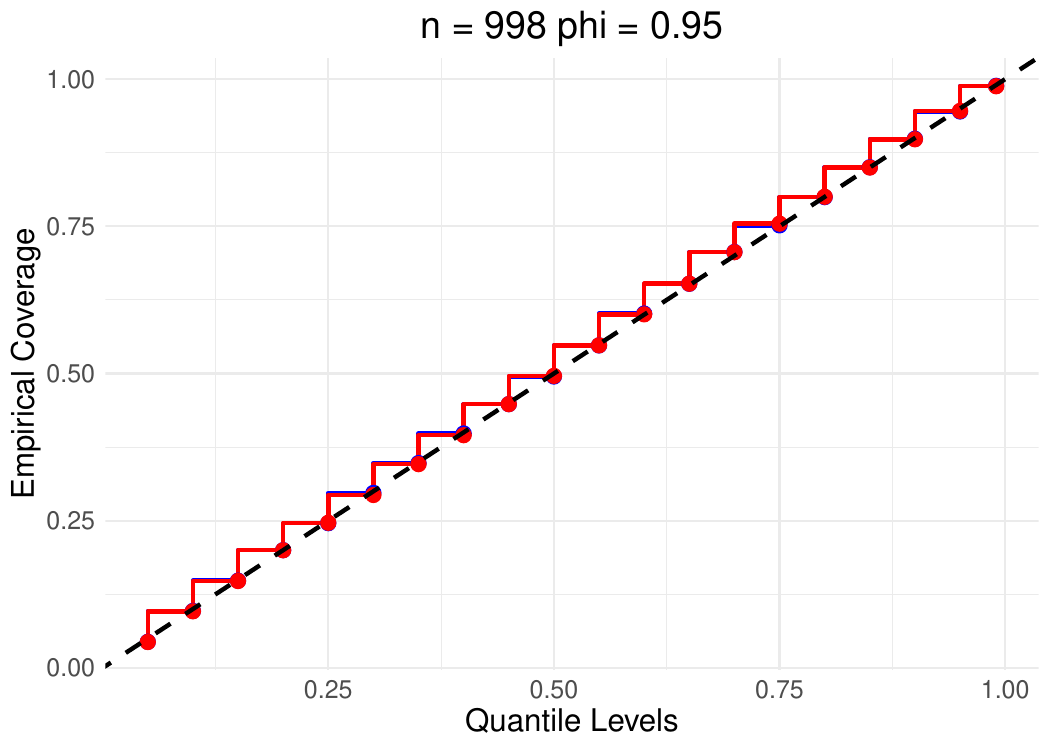}
    }\\
    \vspace{1em}
    \subfloat[\label{}]{       \includegraphics[width=0.29\textwidth]{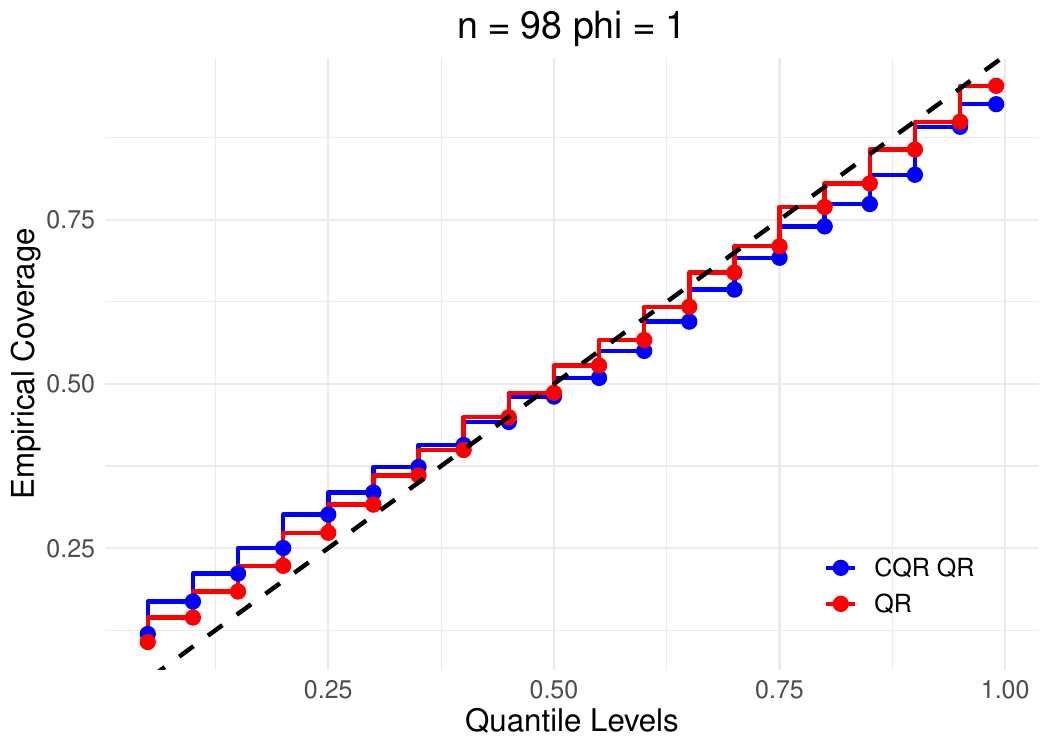}
    }
    \hspace{1em}
    \subfloat[\label{}]{        \includegraphics[width=0.29\textwidth]{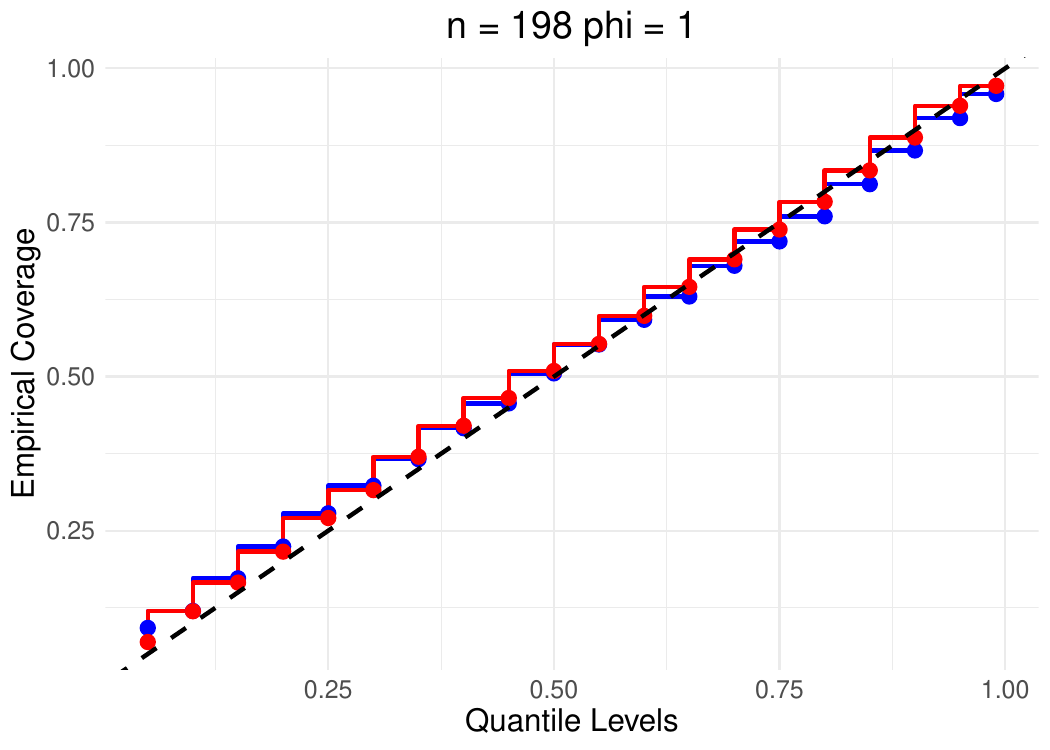}
    }
    \hspace{1em}
    \subfloat[\label{}]{        \includegraphics[width=0.29\textwidth]{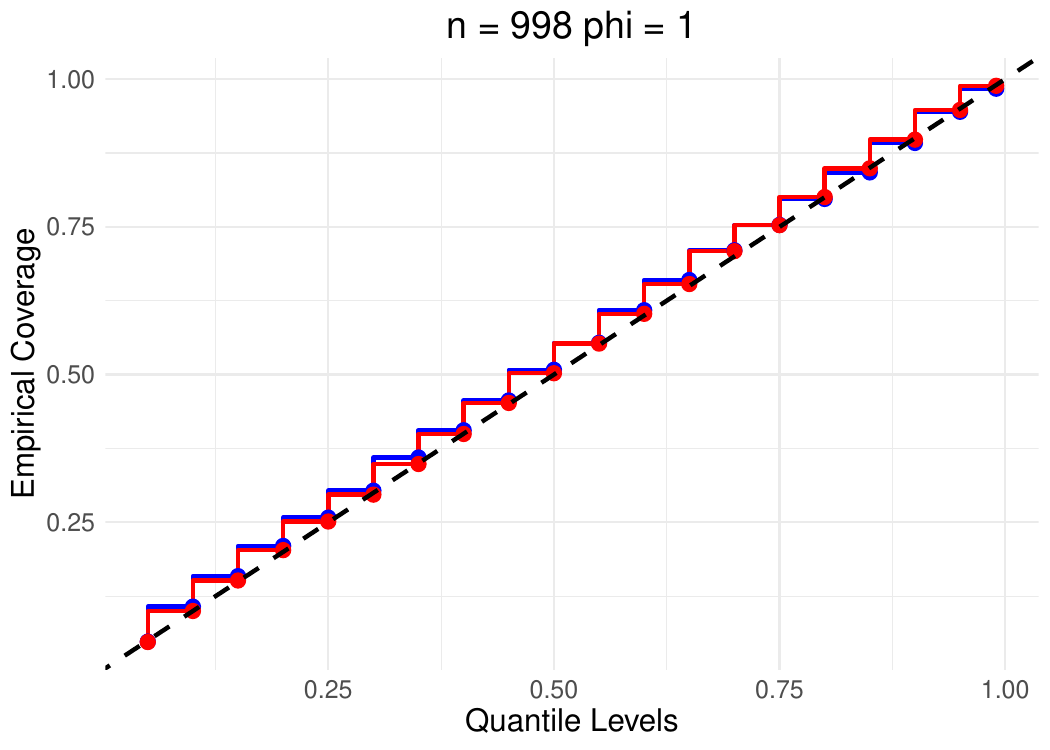}
    }
    \\[10pt]
    \captionsetup{font=footnotesize}
    \caption[]{Calibration curves for QR and CQR QR for the 3 values of \(n\) and \(\phi_1\) = 0.95 (first row), 1 (second row). Results for $\phi_1 = 1.05$ are not reported because both models completely fail to estimate the DGP.}
    \label{fig:figure14}
\end{figure}

\begin{figure}[H]
    \centering
    \subfloat[\label{}]{        \includegraphics[width=0.29\textwidth]{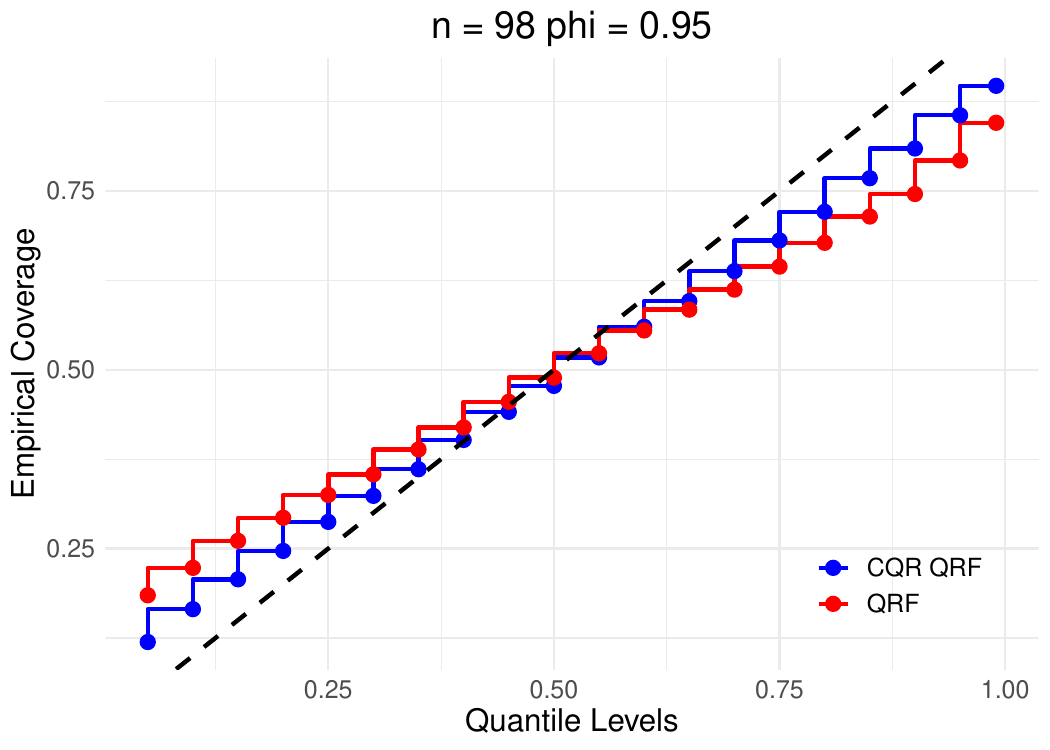}
    }
    \hspace{1em}
    \subfloat[\label{}]{        \includegraphics[width=0.29\textwidth]{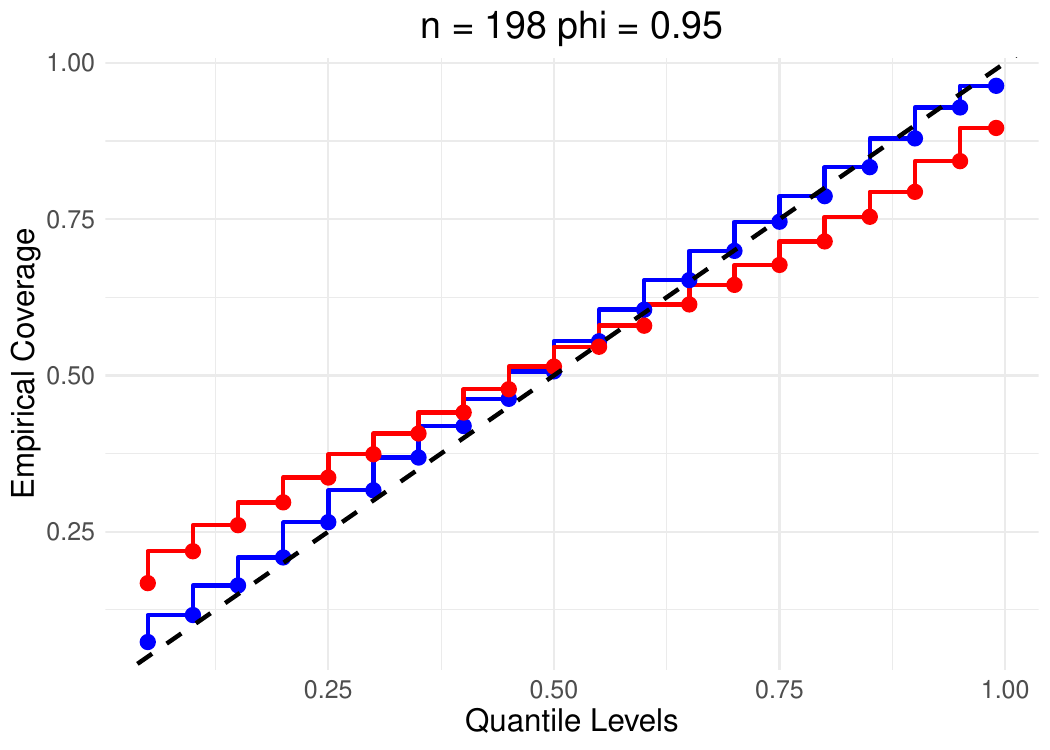}
    }
    \hspace{1em}
    \subfloat[\label{}]{        \includegraphics[width=0.29\textwidth]{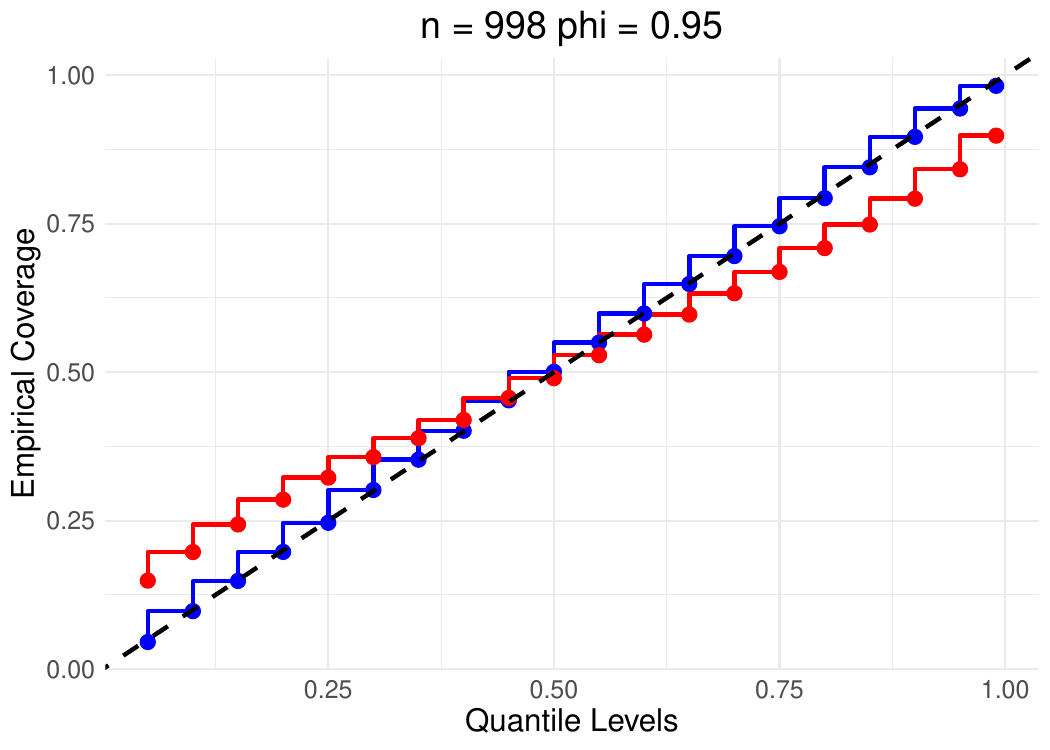}
    }\\
    \vspace{1em}
    \subfloat[\label{}]{       \includegraphics[width=0.29\textwidth]{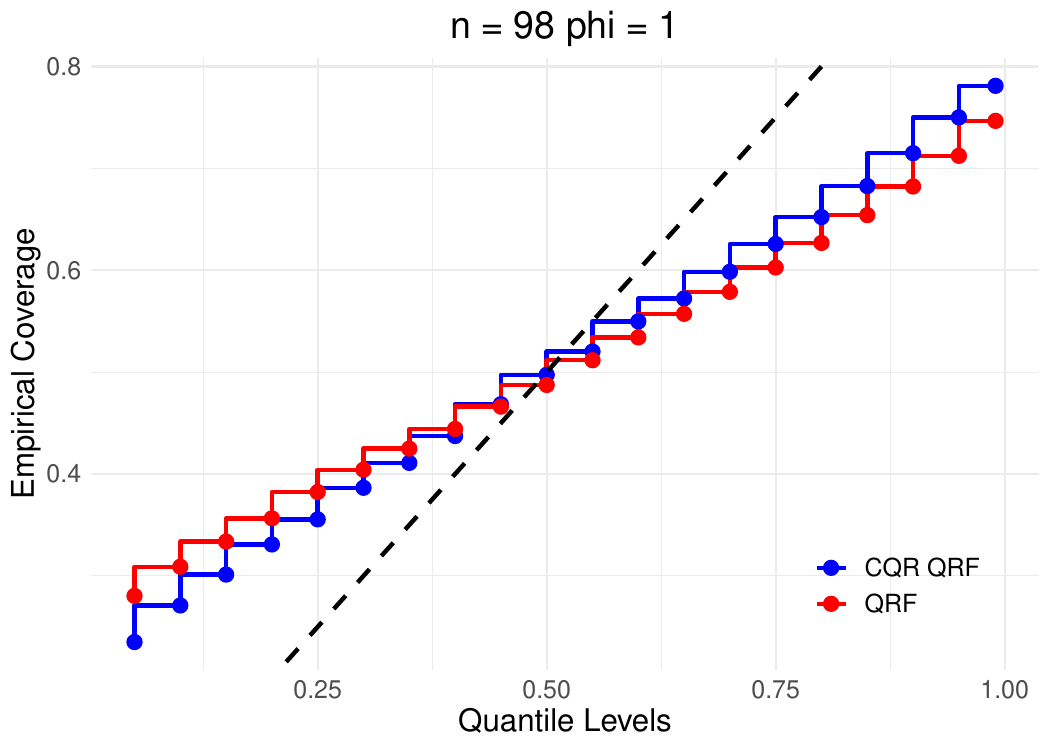}
    }
    \hspace{1em}
    \subfloat[\label{}]{        \includegraphics[width=0.29\textwidth]{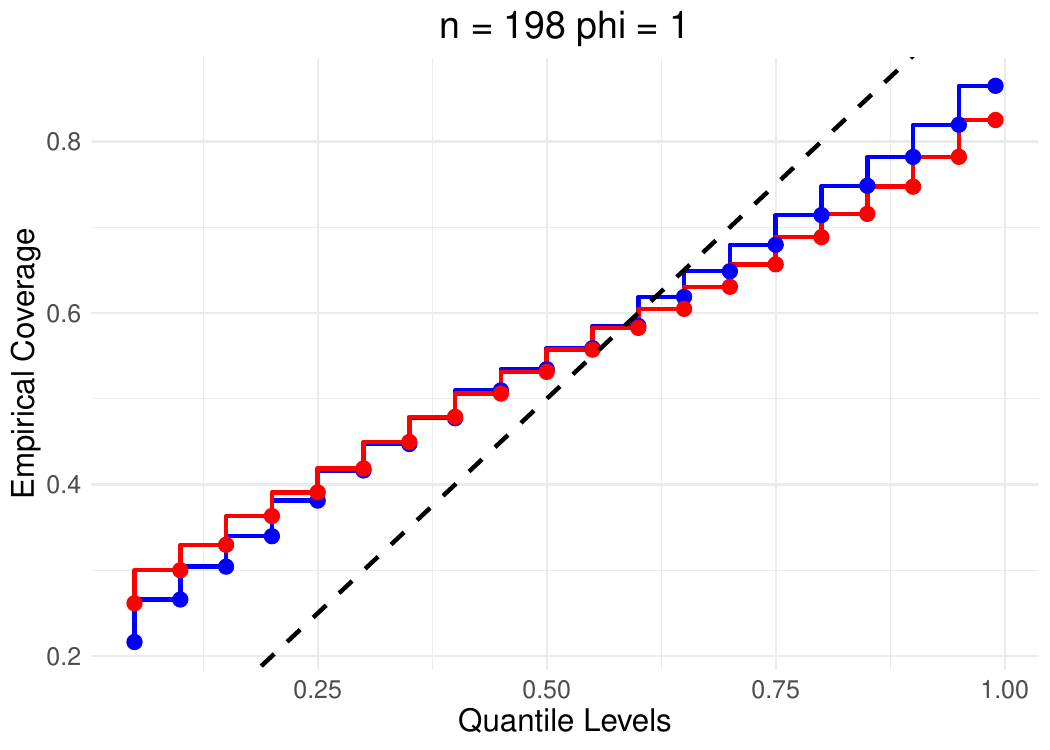}
    }
    \hspace{1em}
    \subfloat[\label{}]{        \includegraphics[width=0.29\textwidth]{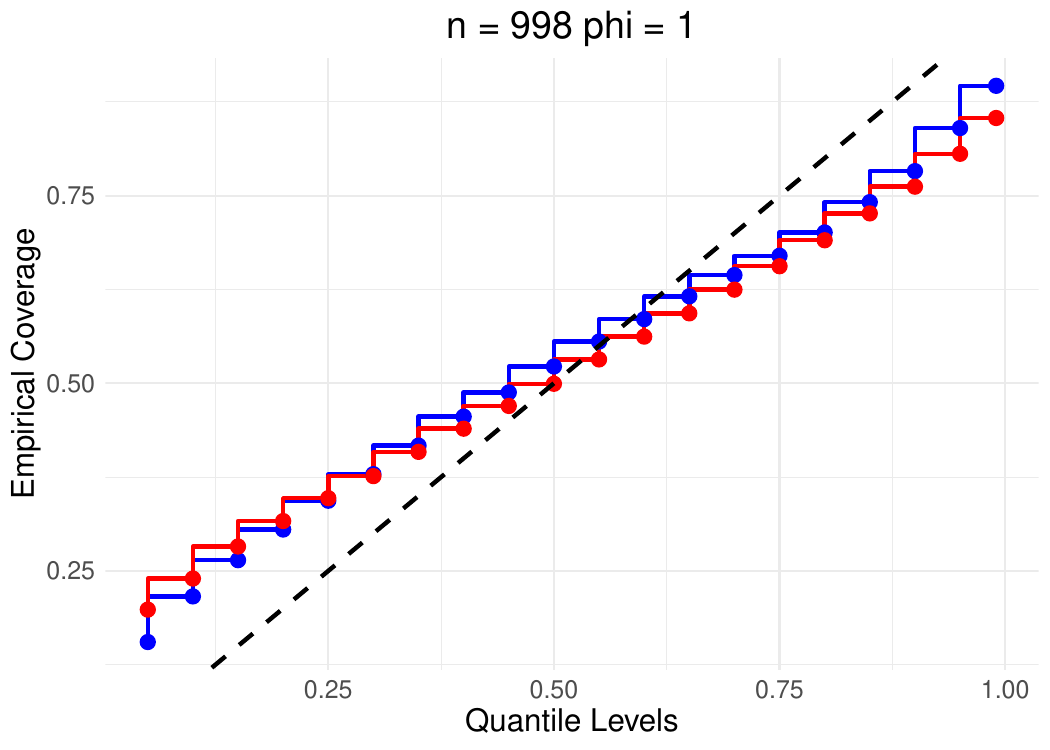}
    }\\[10pt]
    \captionsetup{font=footnotesize}
    \caption[]{Calibration curves for QRF and CQR QRF for the 3 values of \(n\) and \(\phi_1\) = 0.95 (first row), 1 (second row). Results for $\phi_1 = 1.05$ are not reported because all models completely fail to estimate the DGP.}
    \label{fig:figure15}
\end{figure}

From Table \ref{table:table9} and Figures \ref{fig:figure14} and \ref{fig:figure15}, we observe distinct differences in the behaviour of the QR and QRF models. The QR model achieves the highest calibration performance among all models, and its conformalised counterpart (CQR QR) does not provide any further improvement in this regard; in fact, it results in a slight decrease in performance. In contrast, the QRF model faces greater challenges in calibration but shows improvement with the application of conformalisation.
All models improves as the number of observations \(n\) increases.
For the most difficult case, where $\phi_1 = 1.05$, none of the four models successfully captures the dynamics of the data generating process.
In conclusion, conformalisation does not appear a very effective solution for dealing with nearly unit root autoregressive time series. 

\begin{table}[H]
    \centering 
    \begin{tabular}{|p{10em} c c|}
    \hline
    \rowcolor{lavender}
      \textbf{MAE} & \textbf{0.95} & \textbf{1}  \\
    \hline \hline
    \textbf{QR} & 0.007  & 0.015  \\
    \textbf{QRF} & 0.075  & 0.112  \\
    \textbf{CQR QR} & 0.012 & 0.026  \\
    \textbf{CQR QRF} & 0.023  & 0.094   \\
    \hline
    \end{tabular}
    \\[10pt]
    \captionsetup{font=footnotesize}
    \caption{MAE (\ref{eq:MAE}) for the four models with $\phi_1 \in \{0.95, 1\}$, averaged across three different values of \(n\). Results for $\phi_1 = 1.05$ are not reported because both models completely fail to estimate the DGP.}
    \label{table:table9} 
\end{table}


\end{document}